\documentclass{aa}

\usepackage{graphicx}

\usepackage{txfonts}

\usepackage{comment}
\usepackage{ulem}
\usepackage{booktabs}
\usepackage{longtable}
\usepackage{multicol}
\usepackage{caption}
\usepackage{lipsum} 
\usepackage{amsmath}
\usepackage{amssymb}
\usepackage{layout}
\usepackage[algo2e, ruled, vlined]{algorithm2e} 

\usepackage{makecell}
\usepackage{algorithm}
\usepackage{algorithmic}
\usepackage{rotating}
\usepackage{tabularx}

\usepackage{geometry}
\usepackage{natbib,twoopt}
\usepackage[breaklinks=true]{hyperref} 
\usepackage{xcolor}
\bibpunct{(}{)}{;}{a}{}{,}            
\makeatletter
  \newcommandtwoopt{\citeads}[3][][]{\href{http://adsabs.harvard.edu/abs/#3}%
    {\def\hyper@linkstart##1##2{}%
     \let\hyper@linkend\@empty\citealp[#1][#2]{#3}}}
  \newcommandtwoopt{\citepads}[3][][]{\href{http://adsabs.harvard.edu/abs/#3}%
    {\def\hyper@linkstart##1##2{}%
     \let\hyper@linkend\@empty\citep[#1][#2]{#3}}}
  \newcommandtwoopt{\citetads}[3][][]{\href{http://adsabs.harvard.edu/abs/#3}%
    {\def\hyper@linkstart##1##2{}%
     \let\hyper@linkend\@empty\citet[#1][#2]{#3}}}
  \newcommandtwoopt{\citeyearads}[3][][]%
    {\href{http://adsabs.harvard.edu/abs/#3}
    {\def\hyper@linkstart##1##2{}%
     \let\hyper@linkend\@empty\citeyear[#1][#2]{#3}}}
\makeatother

\begin{document} 

    \title{Validating an X-ray-based method for identifying Fermi blazars}

   \subtitle{}

\author{
    J. L. Weiss\inst{1,2,3}\thanks{ \email{joaquin.weiss@inaf.it}}
    \and
    E. J. Marchesini\inst{3}
    \and
    I. Andruchow\inst{1,2}
    \and
    N. Masetti\inst{3,4}
}
   \institute{Facultad de Ciencias Astron\'omicas y Geof{\'\i}sicas  (FCAG), Universidad Nacional de La Plata, Paseo del Bosque s/n, FWA, B1900, La Plata, Provincia de Buenos Aires, Argentina
         \and
             Instituto Argentino de Radioastronom\'ia, CONICET--CICPBA--UNLP, CC5 (1894), Villa Elisa, Provincia de Buenos Aires, Argentina
          \and 
              Osservatorio di Astrofisica e Scienza dello Spazio di Bologna, Istituto Nazionale Di Astrofisica (OASBo-INAF), Via Gobetti 101, I-40129, Bologna, Italy
          \and
                Instituto de Astrof{\'\i}sica, Universidad Andres Bello, Av. Fernandez Concha 700, Las Condes, Santiago, Chile             
        }

   \date{}

  \abstract
   {Blazars are the main population of active galactic nuclei (AGNs) detected in $\gamma$ rays by the Fermi satellite. However, $\sim$34\% of the sources in the Fermi catalog remain unidentified due to relatively large positional uncertainties (approx. 10 arcmin). This means that each $\gamma$-ray detection is positionally correlated with several optical sources, which drives the development of multiwavelength methods, particularly in the infrared and radio, to identify their corresponding counterparts.}
   {A method has been proposed to identify BL Lacertae (BL Lac) candidates among unidentified gamma sources based on their X-ray properties, combining Swift-XRT observations with Fermi source associations. In this work, we aim to validate the effectiveness of this approach.}
   {We empirically assessed the effectiveness of this method for identifying counterparts to Fermi sources by leveraging recently published optical spectroscopic data. From an initial sample of 148 unidentified $\gamma$-ray sources selected as BL Lac candidates based on their X-ray properties, we find that 71 have optical spectra reported in the literature that are spatially consistent with the proposed associations. We further evaluated their classification as BL Lac objects using standard blazar classification criteria. We present a comparative analysis between the initial predictions of the method and the current classifications, highlighting its efficiency in selecting BL Lacs.}
   {Out of the 71 optical counterparts identified for the unidentified Fermi-LAT sources, 80.28\% are confirmed as BL Lac objects and 87.32\% as members of possible $\gamma$-ray-emitting AGNs. These results confirm the high efficiency of the X-ray-based method in selecting blazars and suggest that many of the still unassociated $\gamma$-ray sources are likely to belong to this class. Among the 71 sources, we also identify four Seyfert galaxies. This suggests that the method is likewise sensitive to this class of objects, despite their comparatively rare detection as $\gamma$-ray emitters.}
   {This study confirms that the X-ray band is an effective tool for identifying BL Lac objects among unclassified $\gamma$-ray sources and supports future spectroscopic follow-up campaigns. The high fraction of BL Lac and AGN classifications among the spectroscopically confirmed counterparts suggests that a large portion of the unassociated $\gamma$-ray population is likely dominated by blazars. Interestingly, the identification of a small number of Seyfert galaxies raises an open question in $\gamma$-ray astrophysics regarding the origin of the high-energy emission in systems not typically associated with relativistic jets. This may point to a more complex phenomenology in the low-luminosity AGN population and highlights the need for further multiwavelength follow up.}


   \maketitle

\section{Introduction}

Active galactic nuclei (AGNs) are the most luminous stable sources in the sky with bolometric luminosities on the order of $10^{42}$\,erg\,s$^{-1}$, potentially reaching values up to $10^{48}$\,erg\,s$^{-1}$ \citep{fan}. They can be detected across the entire electromagnetic spectrum \citep{peterson}, showing flux variability on timescales ranging from minutes to years \citep[e.g.,][]{raiteri2025, weiss2026} at all wavelengths, and display high and variable degrees of polarization \citep[e.g.,][]{andruchow2005,pavlidou2014}. By characterizing their broadband emission, it is possible to infer the physical conditions responsible for the radiation generated at the source.

According to the standard and unified models of AGNs \citep{lynden, antonucci1993, urrypadovani}, the observed phenomenology is determined by both the intrinsic structure of the central engine and the observer’s line of sight. In this framework, an AGN consists of a rotating supermassive black hole (SMBH) surrounded by an accretion disk, with fast-moving gas clouds forming the broad-line region (BLR) closer to the nucleus and more distant material constituting the narrow-line region (NLR). In addition, a relativistic jet, launched perpendicular to the accretion disk, is present in many AGNs. Depending on the viewing angle regarding this jet, different observational classes are distinguished. When the jet is closely aligned with the observer’s line of sight, the AGN is identified as a blazar.

Blazars are characterized by an optical spectrum that is largely featureless, typically dominated by a blue continuum, with emission lines either absent or weak. Their spectral energy distribution (SED) displays two broad, nonthermal components. The first is a low-energy bump, generally attributed to synchrotron emission from relativistic electrons in the jet and extending from radio to X-ray frequencies. The second is a high-energy bump, typically associated with inverse Compton scattering \citep[e.g.,][]{ghisellini1985, marscher, bloom, chiadberge, paiano} or hadronic processes \citep[e.g.,][]{dermer, sikora, blazejowski, ghisellini2009} spanning from X-rays to $\gamma$ rays and, in some cases, reaching energies above the TeV regime. These sources are further subclassified according to the rest-frame equivalent width of their optical emission lines. Objects with weak or undetectable emission lines (rest-frame equivalent width $\leq$ 5 Å) are classified as BL Lacertae (BL Lac) objects, whereas those with prominent broad emission lines (> 5 Å) are classified as flat-spectrum radio quasars (FSRQs; \citealt{Laurent, marcha1996}). In general, BL Lacs exhibit lower bolometric luminosities, while FSRQs are intrinsically more luminous sources \citep{mondal2019}. BL Lacs are also subdivided according to the ratio between their radio and X-ray fluxes. In this context, two subclasses distinguish sources on the basis of this ratio \citep{maselli2010}: high-energy peaked BL Lacs (HBLs) and low-energy peaked BL Lacs (LBLs). This classification is commonly expressed through the parameter $
\Phi_{XR} = 10^{-3} \frac{F_X}{S_{1.4}\,\Delta \nu}$, where $F_X$ is the X-ray flux in the 0.1-2 keV band measured by the ROSAT survey \citep{voges1999} and $S_{1.4}\Delta \nu$ is the radio flux density at 1.4 GHz multiplied by the corresponding bandwidth $\Delta \nu$, typically 1 GHz. In this scheme, sources with $\Phi_{XR} > 1$ are classified as HBLs, while those with $\Phi_{XR} < 1$ correspond to LBLs. Some authors also include an intermediate class of BL Lacs (IBLs) \citep[e.g.,][]{maselli2010}. While the total number of known blazars is on the order of 4.000 sources \footnote{4FGL-DR4 catalog: \url{https://heasarc.gsfc.nasa.gov/w3browse/fermi/fermilpsc.html}}, they become the dominant population when observing the sky at $\gamma$-ray energies.

The Large Area Telescope (LAT), on board the Fermi satellite \citep{atwood}, is a space-based instrument that has been surveying the $\gamma$-ray sky since 2008, producing the most comprehensive high-energy source catalog to date (4FGL-DR4; \citealp{abdollahi}). Among the associated sources (that is, those with a reliably identified optical counterpart), $\sim$32\% are classified as BL Lacs, which constitute the majority of extragalactic objects detected at high energies. Nevertheless, 2830 ($\sim$34\%) of the detected sources remain unassociated, motivating the development of multiwavelength strategies to identify their counterparts. The positional uncertainty of Fermi-LAT, typically spanning several arc minutes, makes associating its sources with counterparts at other wavelengths a challenging task. Within this error region, a single $\gamma$-ray source can have hundreds or even thousands of potential counterparts in the optical, radio, infrared, or X-ray bands, making it necessary to establish clear criteria for reliable associations.

The X-ray band provides a particularly powerful diagnostic for blazar identification because it often probes the transition region between the synchrotron and inverse Compton components of the SED. Depending on the source class, the X-ray emission may be dominated by synchrotron radiation (high synchrotron-peaked blazar; HSP), inverse Compton emission (low synchrotron-peaked blazar; LSP), or a combination of both near the crossover between the two components (intermediate synchrotron-peaked blazar; ISP). In particular, X-ray observations have been widely used to search for counterparts of $\gamma$-ray sources  \citep{paggi}. The Swift-XRT telescope\footnote{https://www.swift.ac.uk/LSXPS/} is of particular interest for associating Fermi-LAT sources, as it provides automatic, though irregular, follow-up of Fermi detections \citep{stroh}. XRT observations provide positions with an accuracy on the order of 6 arc seconds. This precision, being several orders of magnitude better than the typical positional error of Fermi-LAT, significantly narrows the search area for optical counterparts.

In the work by \cite{marchesini2019}, a uniform sample of X-ray counterparts of known BL Lac objects based on Swift-XRT observations, was constructed as a basis for studying the relationship between X-ray and $\gamma$-ray emission detected with Fermi-LAT, considering X-ray fluxes and hardness ratios ($\mathrm{HR_x}$). Significantly, the X-ray data were built by co-adding individual snapshots to ensure a minimum effective exposure time of 1 ksec per source, thereby reducing selection biases associated with nonuniform exposure times typical of X-ray observations. Using this sample, they developed a method to classify $\gamma$-ray sources based on their X-ray properties. Building on this approach, they extended the analysis to unassociated Fermi sources \citep{marchesini2020}, identifying candidate BL Lac counterparts.

Blazar classification campaigns have been carried out by various research groups using different strategies, including machine learning approaches \citep[e.g.,][]{agarwal2023, bhatta2024, gharat2024}. More recently, X-ray information has also been incorporated into the search for Fermi counterparts by cross matching with X-ray catalogs and using X-ray properties to classify candidate counterparts \citep{paggi26, neumann26}, but the most robust identification criterion remains the optical spectrum of the source (see, for instance, \citealp{mesetti}). Since the publication of the method, several groups have independently classified Fermi sources based on their optical spectra \cite[e.g.,][]{massaro2016,paiano2017,pena2017}, making it now possible to perform a statistical validation of the method. 

In the present study, we proceeded with the identification of BL Lac candidates, testing the validity of this method by surveying the literature for the optical spectra of sources listed as firm candidates by \cite{marchesini2019}. In Sect. 2, we describe our methodology. In Sect. 3, we present our results and highlight peculiar objects. In Sect. 4, we discuss our results, and in Sect. 5, we summarize our findings and present our conclusions.

\section{Methodology}

\begin{figure}[h!]
    \centering
    \includegraphics[width=0.95\columnwidth]{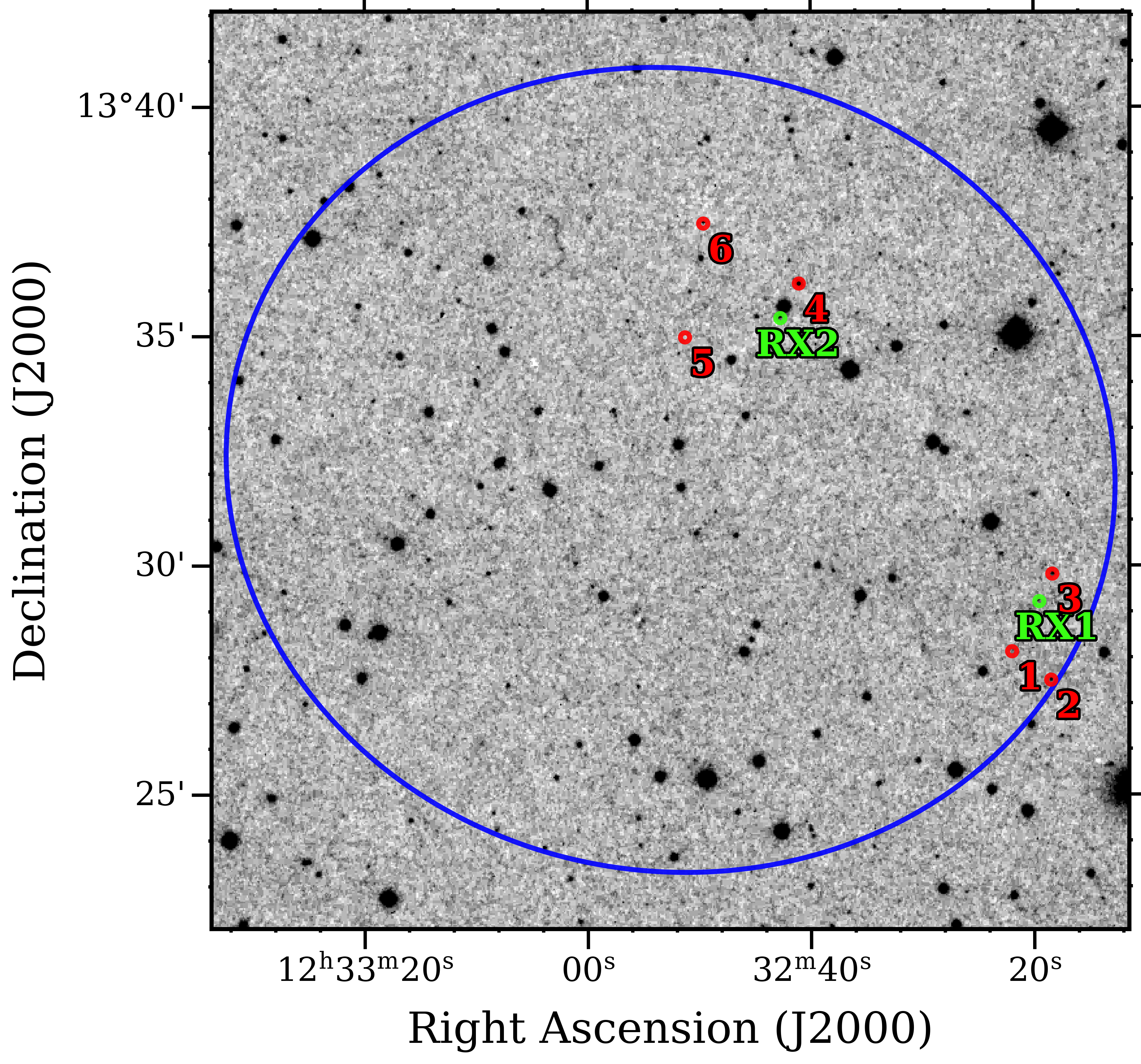}
    \caption{Image of the source 3FGL J1232.8+1332 taken from the Digitized Sky Survey (DSS) \citep{lasker1994}. The Fermi error ellipse is shown in blue; the positional uncertainty of the X-ray counterparts, selected according to the criterion of \cite{marchesini2020}, is shown in green; and the optical counterparts with available spectra located near the X-ray detections is shown in red.}
    \label{fig:ds9}
\end{figure}

We began from a sample of 148 unidentified $\gamma$-ray sources (UGSs) from the 3FGL catalog. These sources were selected for having at least one X-ray detection, obtained with Swift-XRT, within the 95\% Fermi-LAT error ellipse, following the selection and analysis criteria described in \cite{marchesini2020}, namely Galactic latitude $|b|>10^{\circ}$, exposure time between 1 and 10 ksec, and signal to noise ratio (S/N) of the detection greater than 3. As an example, Fig. \ref{fig:ds9} shows the Fermi error ellipse of the source 3FGL J1232.8+1332, in contrast with the observational uncertainties from Swift and the image of the field. It is important to note that this initial selection of 148 candidates was performed based solely on their X-ray and $\gamma$-ray properties, without prior knowledge of the existence of optical spectroscopic data. The literature search was carried out subsequently preventing selection bias in the evaluation of the method's efficiency.

From this sample, we carried out a search for possible optical counterparts in the literature. As a starting point, a crossmatch was performed using the \textit{TOPCAT}\footnote{\url{http://www.star.bris.ac.uk/~mbt/topcat/}} software, employing the coordinates of the X-ray detections obtained with Swift and considering a positional uncertainty of 6 arc seconds as adopted by \cite{marchesini2020} in their counterpart identification procedure. The crossmatch was conducted with the optical catalogs in \textit{Simbad} \citep{wenger} and \textit{Aladin} \citep{bonnarel}. The identified sources were, in all cases, associated through machine-learning based criteria, which we did not consider conclusive due to the lack of spectroscopic confirmation. We excluded these associations from our validation sample because optical spectroscopy remains the most widely accepted criterion for robust blazar classification. Despite this, taking this classification into account, the crossmatch returned 123 sources with catalog classifications, of which 75 are listed as blazars (61.0\%) and 82 as AGNs (66.6\%), while the remaining objects belong to source classes not relevant for the present work.

As a following step, we conducted an extensive search in the literature for those sources classified based on optical spectra, using data from peer-reviewed publications, the Sloan Digital Sky Survey spectral catalog (SDSS) - DR19\citep{sdss} and the Large Sky Area Multi-Object Fiber Spectroscopic Telescope (LAMOST) survey - DR9\citep{cui2012}\footnote{LAMOST Data Release 9; \url{http://www.lamost.org/dr9/}}. The corresponding associations were found distributed across 19 previous studies: \cite{shaw, landoni, alvarez, massaro2016, marchesini2016, paiano2017, pena2017, linares, marchesi2018, desai2019, paiano2019, pena2020, rakshit, pena2021, rajagopal, sexton, garcia2023, xu, ulgiati}. 

We also performed a cross-match with the Wide-field Infrared Survey Explorer (WISE) catalog \citep{wise} to investigate the location of the sources in the infrared color–color diagram. This allows us to assess whether the identified objects occupy the region typically associated with blazars, which are known to be well separated from other source classes in this parameter space \citep{dabrusco2012}.

Since the different bibliographic sources use their own nomenclatures, in this work the classifications were standardized into BL Lac, FSRQ, quasi-stellar objects (QSOs), Seyfert galaxies (grouping together Seyfert 1, Seyfert 2, and narrow-lined Seyfert 1 classes), and inactive galaxies. It is worth noting that, although the spectra of BL Lac objects are typically featureless (i.e., showing no absorption or emission lines), classifications based only on this characteristic were not considered, since it is not sufficient to confirm the BL Lac nature of an object (see, e.g., \citealp{Laurent}), and were therefore treated as unobserved sources. 

\begin{figure}[h!]
    \centering
    \includegraphics[width=0.95\columnwidth]{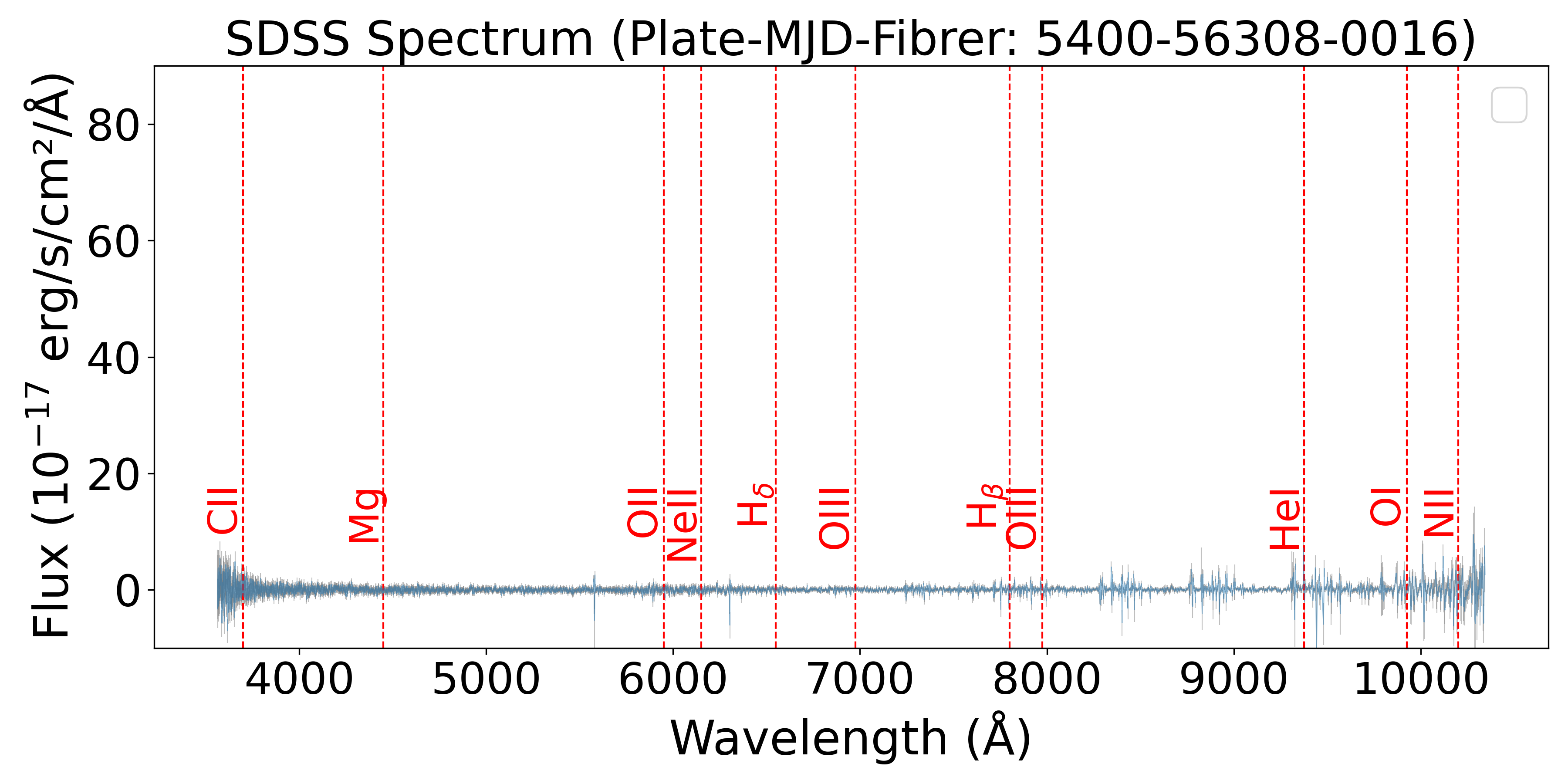}
    \includegraphics[width=0.95\columnwidth]{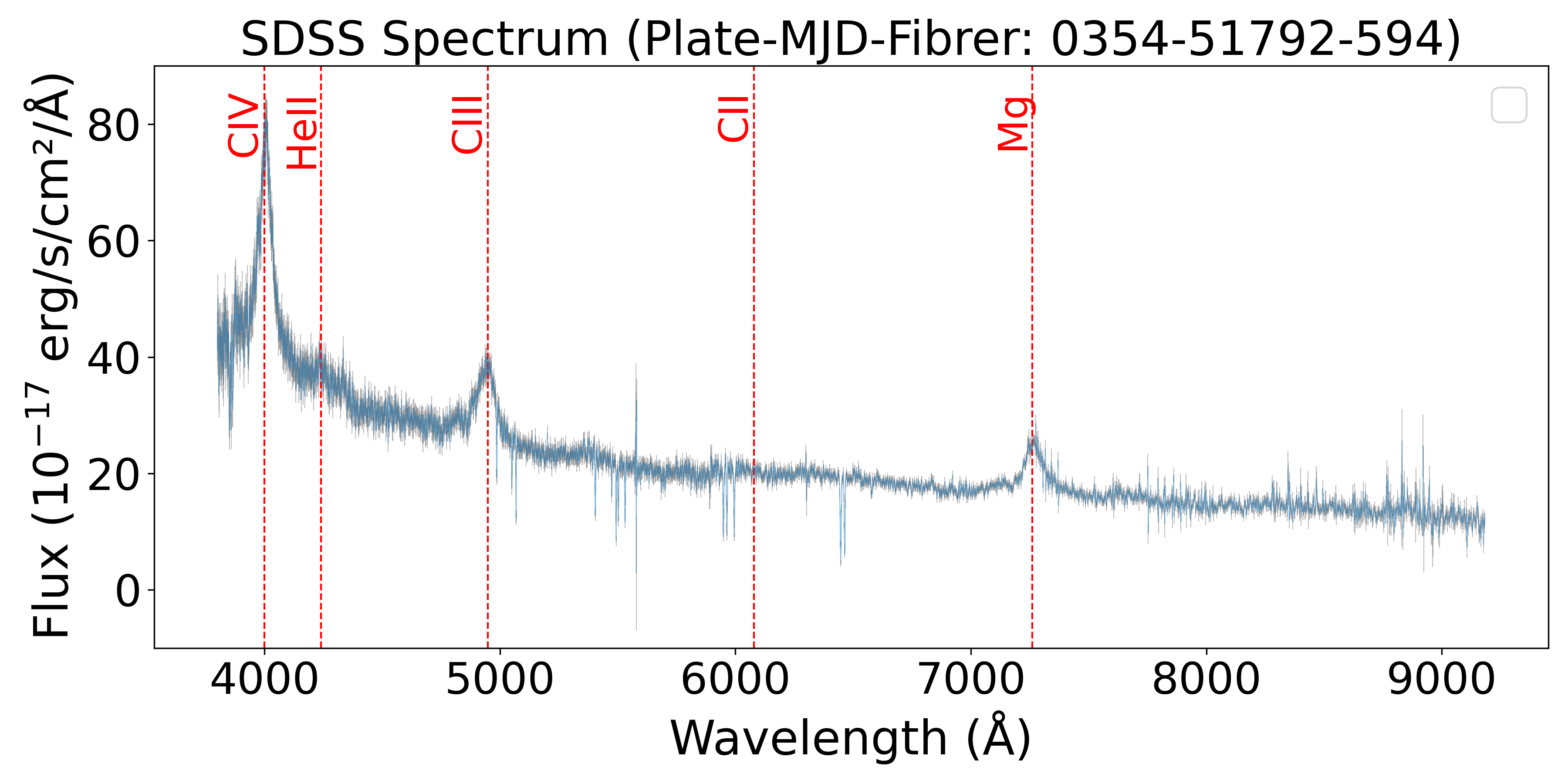}
    \caption{SDSS spectra, both classified as QSOs, with very different signal-to-noise ratios.
\textbf{Top panel:} Spectrum with S/N < 3, where the line association and source classification are unreliable.
\textbf{Bottom panel:} Spectrum with S/N > 3, where the classification is reasonably reliable.
}
    \label{fig:sloan}
\end{figure}

 It is worth noting that SDSS assigns classifications automatically, identifying emission or absorption lines through template fitting (see, e.g., \citealp{bolton}) and, based on this, estimates the redshift of the source, without accounting for the S/N of the spectrum or potential contamination from artifacts (see, for example, Fig.~\ref{fig:sloan}, where two clearly disparate spectra are both classified as QSOs). Spectra flagged in the database, as well as those with a S/N lower than 3, were excluded from the analysis. Each remaining spectrum was inspected and classified following the criteria described in \cite{Laurent}. As a first step, each spectrum was checked to identify spectral lines and assess their reliability. When spectral features were detected, the source redshift was recalculated based on the line identifications. Particular attention was also given to measuring the amplitude of the calcium (Ca II) break, evaluating the presence or absence of spectral lines and their widths, and verifying the existence of radio-band detections for these sources. In cases where counterparts were obtained at two different frequencies, the spectral index of the source was calculated, since at these frequencies the dominant emission in blazars is expected to be synchrotron radiation, for which the spectrum is expected to be flat. For this purpose, the NRAO VLA Sky Survey (NVSS; 1.4 GHz; \citealp{condon}), Faint Images of the Radio Sky at Twenty-Centimeters (FIRST; 1.4 GHz; \citealp{becker}), the Sydney University Molonglo Sky Survey (SUMSS; 843 MHz; \citealp{mauch}) and the TOCats catalog\footnote{\url{
https://catsweb.oas.inaf.it/}} (Nicastro, L. et al. in prep) were used. This inspection and classification process yields a sample of 71 spectroscopically identified sources.

Additionally, we identified seven Fermi sources for which the optical counterpart reported in the literature is not spatially consistent with the X-ray source selected by the procedure of \cite{marchesini2020}. For these cases, we examined both the literature counterpart and the X-ray source identified by our method in order to assess the reliability of the proposed associations.

\begin{figure}[h!]
    \centering
    \includegraphics[width=0.85\columnwidth]{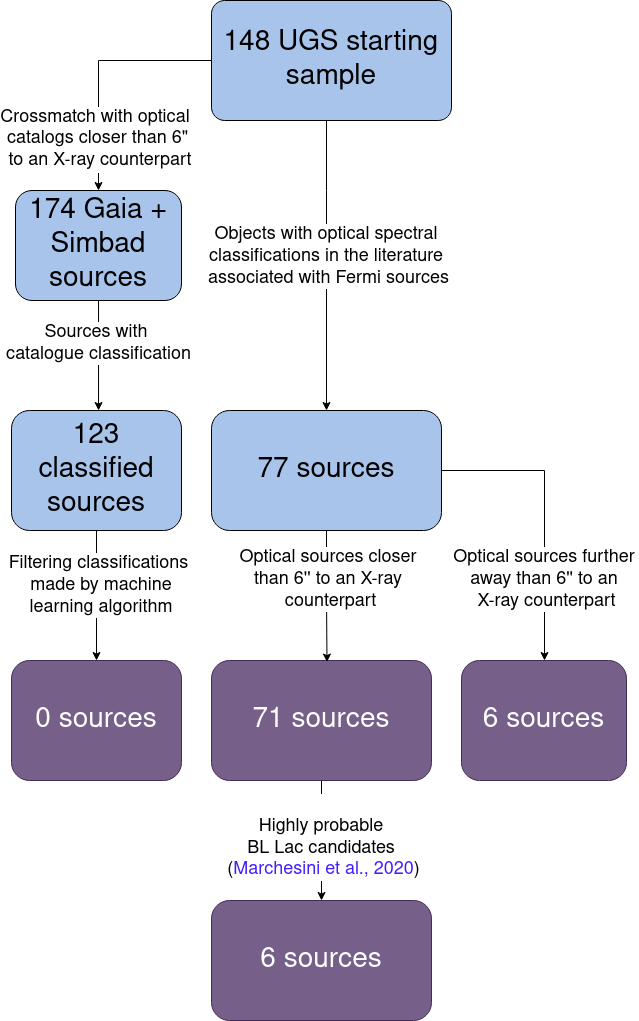}
    \caption{Flowchart summarizing the filtering process of the sources analyzed in this work.}
    \label{fig:flowchart}
\end{figure}

Finally, we examined which of the sources identified in our study matched any of the 32 X-ray sources listed as highly probable BL Lac candidates by \cite{marchesini2020}. Figure \ref{fig:flowchart} outlines the selection steps discussed above.

\section{Results}

In total, 77 UGSs were identified as being associated with a Fermi-LAT source through the acquisition of an optical spectrum of at least one source within its error ellipse. Subsequently, we considered only those cases in which the optical counterpart reported in the literature coincide positionally with the X-ray source. Under this criterion, the sample was reduced to 71 sources, the results are presented in Table~\ref{tab:resultados2}. That table shows the associations found, including the increment factor (IF). We define the IF as the ratio between the percentage of a class in our sample and the percentage reported in the 4FGL-DR4 catalog. It should be noted that this IF is used as a comparison indicator, and not as a formal statistical one.

Among the 71 sources, 80.28\% were ultimately classified as BL Lacs. It is important to emphasize that, since the selection criteria are strongly based on X-ray detections, a high identification rate of BL Lac objects is expected. This is due to their characteristic strong nonthermal emission extending into the X-ray band, in addition to thermal emission from the accretion disk. When possible, $\gamma$-ray-emitting AGNs are considered (BL Lac, FSRQ, and Seyfert galaxies). The method identifies 87.32\% of the sources as belonging to this class. 

\begin{table}[!t]
\centering
\caption{Spectroscopic classification considering only optical spectra obtained within 6 arc seconds of the X-ray source.}
\begin{tabular}{cccccc}
\hline\hline\noalign{\smallskip}
Class & Quantity & \% & \% Fermi & IF\\
\hline
BL Lac & 57 & 80.28 & 32.15 & 2.50 \\
QSO & 7 & 9.86 & - & - \\
Seyfert & 4 & 5.63 & 0.26 & 21.65\\
FSRQ & 1 & 1.41 & 17.69 & 0.08 \\
Galaxy & 1 & 1.41 & 0.13 & 10.85 \\
Star & 1 & 1.41 & - & - \\
\hline
\end{tabular}
\tablefoot{
Column 1 indicates the source class; column 2 lists the number of sources in our sample associated with that class; column 3 gives the corresponding percentage; and column 4 presents the IF of our detections relative to the 4FGL-DR4 catalog.}

\label{tab:resultados2}
\end{table}

Furthermore, the method enables the identification of Seyfert-type active nuclei, whose $\gamma$-ray emission is still not fully understood. In the 4FGL-DR4 catalog, only 12 out of 4635 sources are classified as Seyferts, representing 0.26\% of the total classified sources. In our analysis, out of 71 classified sources, 4 were found to be galaxies of this type. This represents a 5.63\% of the identified sample.

Out of the seven sources classified as QSOs, we further examined their radio emission to verify this classification, since these objects can be misclassified as FSRQs \citep{urrypadovani}. To this end, we followed the radio luminosity ranges reported by \cite{padovani2017} and \cite{ghisellini2017}, and we computed the corresponding radio luminosities of the sources. Radio-weak QSOs (RQs)  are typically characterized by radio luminosities below $\sim10^{25}\,\mathrm{W\,Hz^{-1}}$, while radio-loud (RL) QSOs generally range from $\sim10^{26}$--$10^{28}\,\mathrm{W\,Hz^{-1}}$. Flat-spectrum radio quasars are usually found at radio luminosities between $\sim10^{26}$ and $10^{29}\,\mathrm{W\,Hz^{-1}}$. It is worth noting that the RL QSO and FSRQ populations overlap in the interval $\sim10^{26}$--$10^{28}\,\mathrm{W\,Hz^{-1}}$, where radio luminosity alone cannot provide an unambiguous classification. In such cases, additional information, such as the radio spectral index, can help discriminate between the two classes, since FSRQs are expected to exhibit flatter radio spectra due to the dominant contribution of relativistically beamed jet emission.

Using these thresholds, we find that all sources classified as quasars are consistently confirmed as such, with one exception: for the source associated with 3FGL J0340.4-2423, no radio information could be found in the available catalogs. The results for these seven sources are reported in Table~\ref{tab:qso}. We further find that the remaining objects are consistent with the RQ quasar population. However, we cannot exclude the possibility that some of these associations arise from contamination. While X-ray quasars are relatively common, only a very small fraction are detected as $\gamma$-ray emitters.

\begin{table*}[!t]
\centering
\caption{List of sources classified as QSOs, showing the associated Fermi and Swift source names.}
\begin{tabular}{cccccc}
\hline\hline\noalign{\smallskip}
Name 3FGL & Name XRT &  Flux [mJy] & z & Luminosity [W$\times$ hz$^{-1}$] & Class \\

\hline
3FGLJ0340.4-2423 & SWXRTJ034050.0-242255& - & 0.683 & - & QSO \\
3FGLJ1117.7+0217 & SWXRTJ111731.0+021555& $\leq$ 1.00 & 0.479 & $\leq$ 1.281e+24 & RQ QSO \\
3FGLJ1120.6+0713 & SWXRTJ112042.5+071312& $\leq$ 1.01 & 0.287  & $\leq$ 3.386e+23 & RQ QSO \\
3FGLJ1309.0+0347 & SWXRTJ130832.3+034409& $\leq$ 1.00 & 0.618 & $\leq$ 2.593e+24 & RQ QSO \\
3FGLJ1517.0+2637 & SWXRTJ151650.0+263635& $\leq$ 1.00 & 1.012 & $\leq$ 1.082e+25 & RQ QSO\\
3FGLJ2030.5-1439 & SWXRTJ203028.0-143921& 35.30 & 0.610 & 8.821e+25 & RQ QSO \\
3FGLJ2212.5+0703 & SWXRTJ221231.0+070651& $\leq$ 0.86 & 1.001 & $\leq$ 8.985e+24 & RQ QSO\\
\hline
\end{tabular}
\tablefoot{
The table also lists the radio fluxes at 1.4 GHz, redshifts, and the resulting reclassification based on the adopted criteria.}
\label{tab:qso}
\end{table*}

Subsequently, we examined the subset of 32 X-ray sources listed as highly probable BL Lac candidates by \cite{marchesini2020}. Within our second sample (optical spectra within 6 arc seconds of the X-ray detection), only six of these sources are found to have optical spectra reported in the literature. The properties of these sources are summarized in Table~\ref{tab:resultados3}. All these sources are found to be subclasses of AGNs. Four out of the six are BL Lac objects, we also find one Seyfert-type galaxy and one RQ QSO.

\begin{table*}[ht!]
\centering
\caption{List of sources with their corresponding Swift X-ray counterparts identified as highly probable BL Lac objects by \cite{marchesini2020}, with available classifications based on optical spectra.}

\begin{tabular}{ccccc}
\hline\noalign{\smallskip}
Name 3FGL & Name XRT & Class & $\Delta$$\theta$ & Reference \\
\hline
3FGLJ0031.6+0938 & SWXRTJ003159.7+093617 & Seyfert& 2.39 & 1 \\
3FGLJ0430.1-3103 & SWXRTJ042958.8-305935 & BL Lac & 1.54 & 2 \\
3FGLJ1249.5-0546 & SWXRTJ124919.6-054539 & BL Lac & 4.02 & 3 \\
3FGLJ1309.0+0347 & SWXRTJ130832.3+034410 & RQ QSO & 4.61 & 4 \\
3FGLJ1810.7+5335 & SWXRTJ181037.9+533502 & BL Lac & 1.22 & 5 \\
3FGLJ1934.2+6002 & SWXRTJ193419.8+600141 & BL Lac & 1.66 & 3 \\
\hline
\end{tabular}
\tablebib{
(1)~\citet{paiano2019}; (2) \citet{pena2021}; (3) \citet{garcia2023}; (4) \citet{ulgiati};
(5) \citet{pena2020}.
}
\tablefoot{
Column 1 lists the 3FGL name, column 2 the Swift counterpart, column 3 the classification, column 4 the angular separation between the Swift and optical positions measured in arc seconds, and column 5 the corresponding reference for the optical spectra.}
\label{tab:resultados3}
\end{table*}

\subsection{Estimation of contamination due to false positives}

A number of Swift-XRT sources is expected to be associated to Fermi fields only by chance, highly depending on the density of detected sources in the sky. Thus, to estimate this rate of false positives in the original sample of 71 XRT counterparts, we used the 2SXPS Swift-XRT point source catalog \citep{Evans20}. This catalog uses a similar method for source detection to that of \cite{marchesini2020}, stacking different snapshots to increase the total exposure time, and lists $>$\,200000 unique sources over $\sim$3800 square degrees. 

We queried the 2SXPS catalog applying the same selection criteria used in our own search, ($|b|>10^{\circ}$, exposure time between 1 and 10 ksec, and S/N of the detection greater than 3). This yielded 17480 unique sources. We then computed the area covered by each dataset (individual observation or stacked image) contributing to this sample, using the reported {\it ImageSize} field. Summing these areas, after counting each field only once to avoid double-counting fields containing multiple sources, we derived a total surveyed area of $6230.7\, \mathrm{deg}^{2}$. Dividing the number of sources by this area gives a background source density of $\rho \sim 2.81 \,\mathrm{deg}^{-2}$. 

We then computed the area of the 95\% confidence error ellipse for each of the 71 Fermi sources in our sample as $\pi$ times the semimajor and semiminor axes reported in the 3FGL catalog and summed these to obtain a total sample area of $1.54\, \mathrm{deg}^{2}$. Multiplying this by the background density yields an expected number of chance coincidences of $N_{FP} \sim \rho \times A_{total} \sim 4.3$, corresponding to a false positive fraction of approximately 6\% of the sample. Given Poisson statistics, we adopt $4 \pm 2$ as our estimate of the number of counterparts in our sample that may be due to chance positional coincidence with an unrelated X-ray source rather than genuine association with the Fermi target.

\subsection{Notes on peculiar objects}\label{sec:peculiar}

The sources presented in this section require individual discussion because their classification could not be established as unambiguously as for the rest of the sample. In these cases, we find discrepancies between previously proposed counterparts and the associations obtained through our analysis. These discrepancies suggest that some literature classifications may require revision or that observational properties do not allow for a unique interpretation. We therefore discuss each source separately and assess the available evidence supporting its proposed nature.

\subsubsection{3FGL J0006.2+0135}\label{sec:J0006}

\begin{figure*}[t!] 
    \centering

    \includegraphics[width=0.45\textwidth]{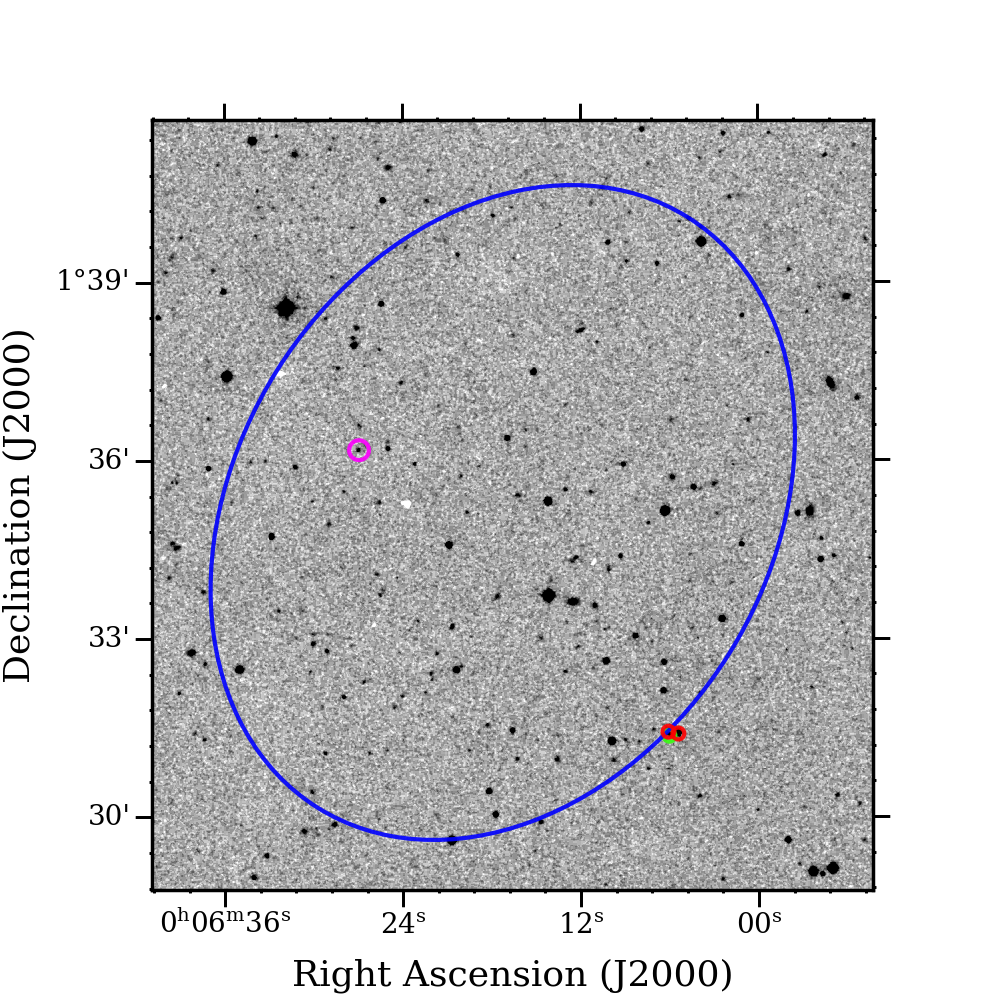}
    \vspace{2mm} 
    \includegraphics[width=0.45\textwidth]{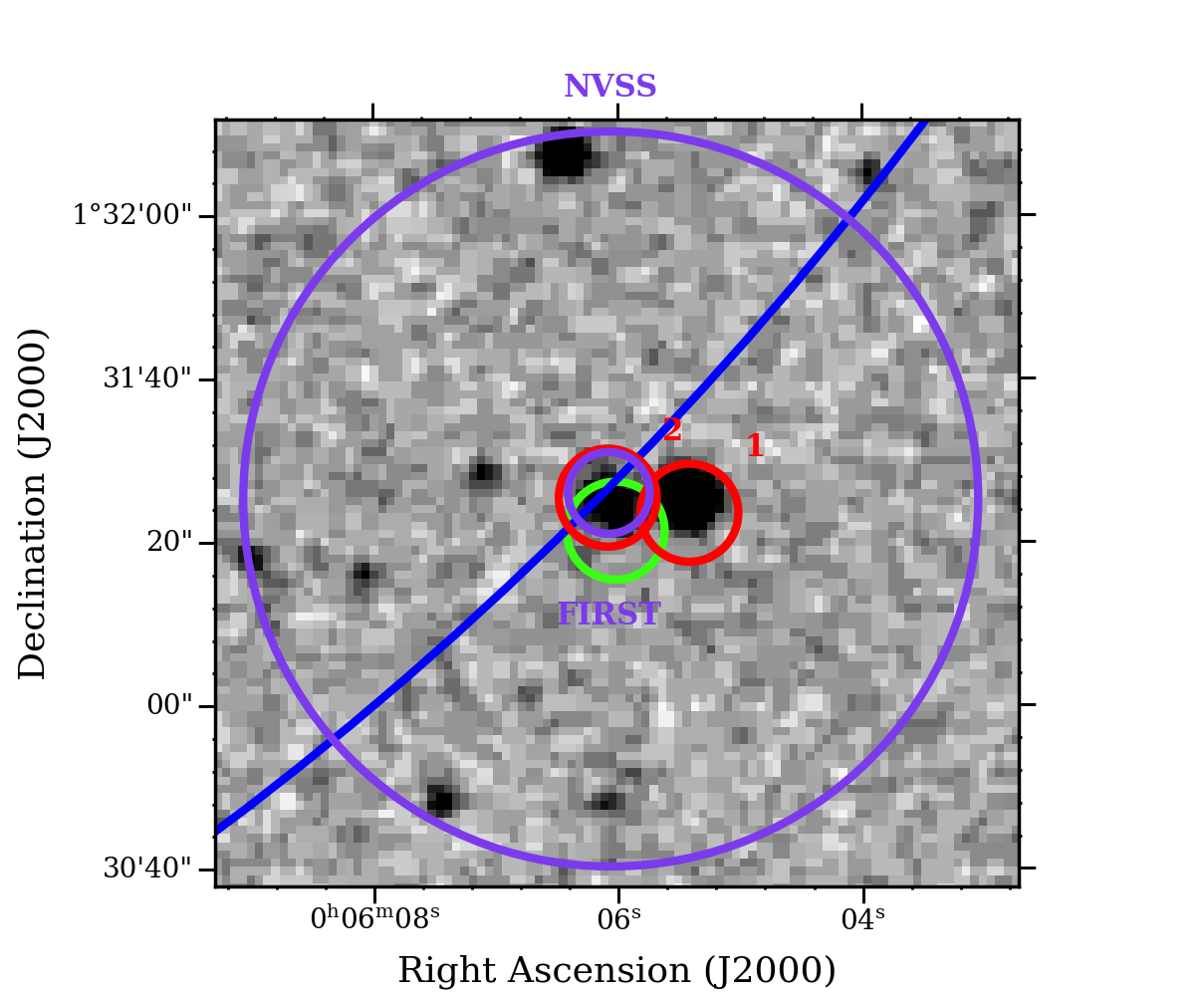}
    
    \caption{Field of the $\gamma$-ray source 3FGL J0006.2+0135, showing the Fermi error ellipse (blue), the X-ray candidate counterpart selected according to our method (green), the counterpart with an optical spectrum available in the literature ( magenta), the optical counterparts located within the Swift positional uncertainty (red). All images were taken from DSS. \textbf{Left panel:} Full optical field for the Fermi detection. \textbf{Right panel:} Zoom-in on the region where the X-ray detection and the optical counterparts labeled "1" and "2" can be seen, including the radio detections with their positional uncertainties in violet.}
    \label{fig:J0006.2+013}
\end{figure*}

This source was classified by \cite{paiano2019} as a BL Lac object, assuming the ROSAT faint source 1RXSJ000626.6+01360 as its X-ray counterpart. However, this position does not spatially coincide with the X-ray counterpart detected proposed by \cite{marchesini2020} (see Fig.~\ref{fig:J0006.2+013}). According to \cite{marchesini2019}, 97\% of Fermi blazars have an X-ray detection with a signal-to-noise ratio (S/N) > 3, suggesting this source could belong to the remaining 3\%. Notably, at the position of the proposed X-ray counterpart, there are two spatially coincident optical sources, labeled as  "1" and "2" in Fig.~\ref{fig:J0006.2+013}. Object 1 (Gaia DR3 2738766508536660480, G=15.43 mag) has an available optical spectrum from SDSS DR16, which classifies it as a galaxy. In contrast, Object 2 (Gaia DR3 2738766538600850176, G=19.56 mag) currently lacks spectroscopic observations. Furthermore, we found coincident radio emission centered on the Swift-XRT counterpart (Source IDs: FIRST J000606.0+013126 and NVSS 000606+013125), with fluxes of 27.80 mJy and 22.82 mJy, respectively. Given the presence of $\gamma$ ray, X-ray, optical, and radio emission, this source is a strong BL Lac candidate and it will be possible to confirm its nature once we obtain its optical spectrum.

\subsubsection{3FGLJ1132.0-4736}\label{sec:J1132}

This object was originally classified as a galaxy by \cite{pena2017}. However, upon inspection of its optical spectrum and following the criteria proposed by \cite{Laurent}, we find no detectable emission lines, a very weak or absent Ca II break and a blue continuum excess at wavelengths shorter than the Ca II break region. In addition, no prominent hydrogen Balmer lines are observed in the spectrum. The source is also associated with a radio counterpart located at a distance of 1.9 arc seconds, with a flux density of 18.4 mJy at 843 MHz. Its infrared colors are consistent with the WISE blazar strip (see section~\ref{sec:disc}). Taken together, these observational properties strongly support a BL Lac classification, and therefore we consider this source to be a BL Lac object throughout this work.

\subsubsection{3FGL J1627.8+3217}\label{sec:J1627}

\begin{figure*}[t!] 
    \centering
    \includegraphics[width=0.49\textwidth]{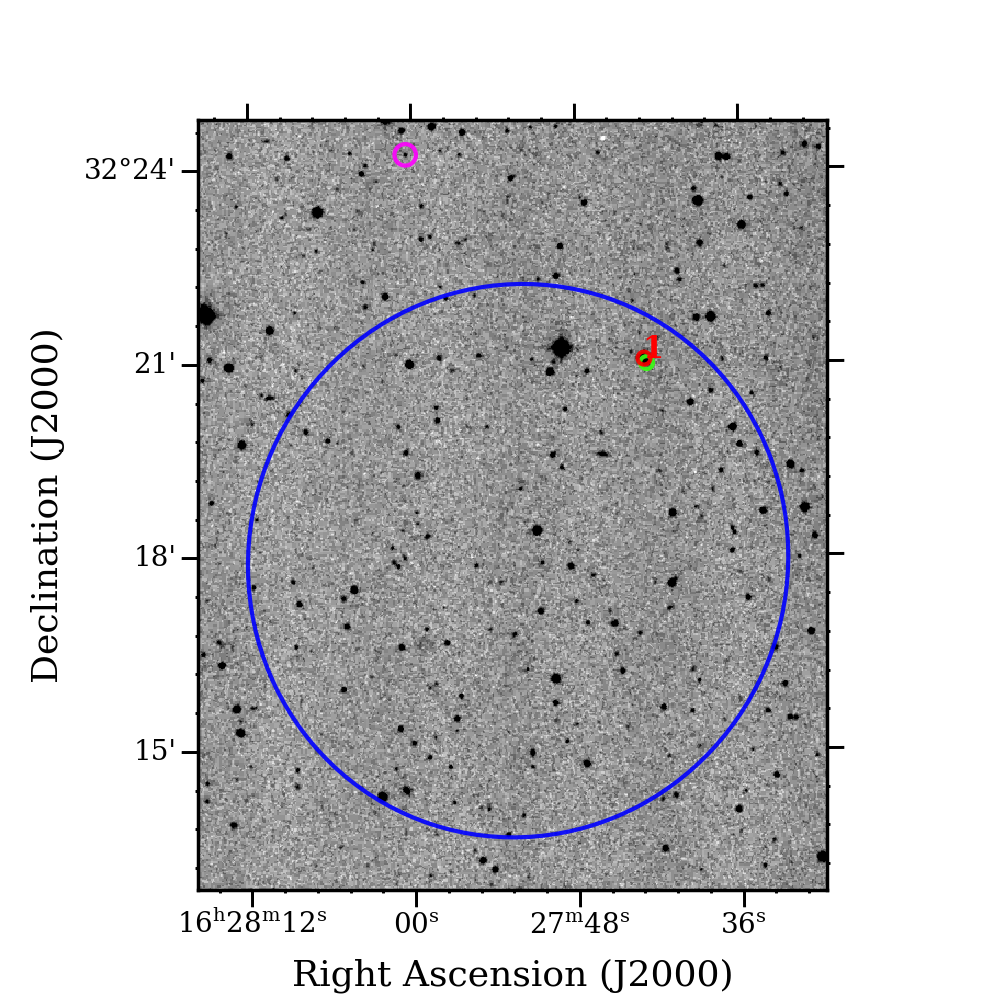}\hfill
    \includegraphics[width=0.49\textwidth]{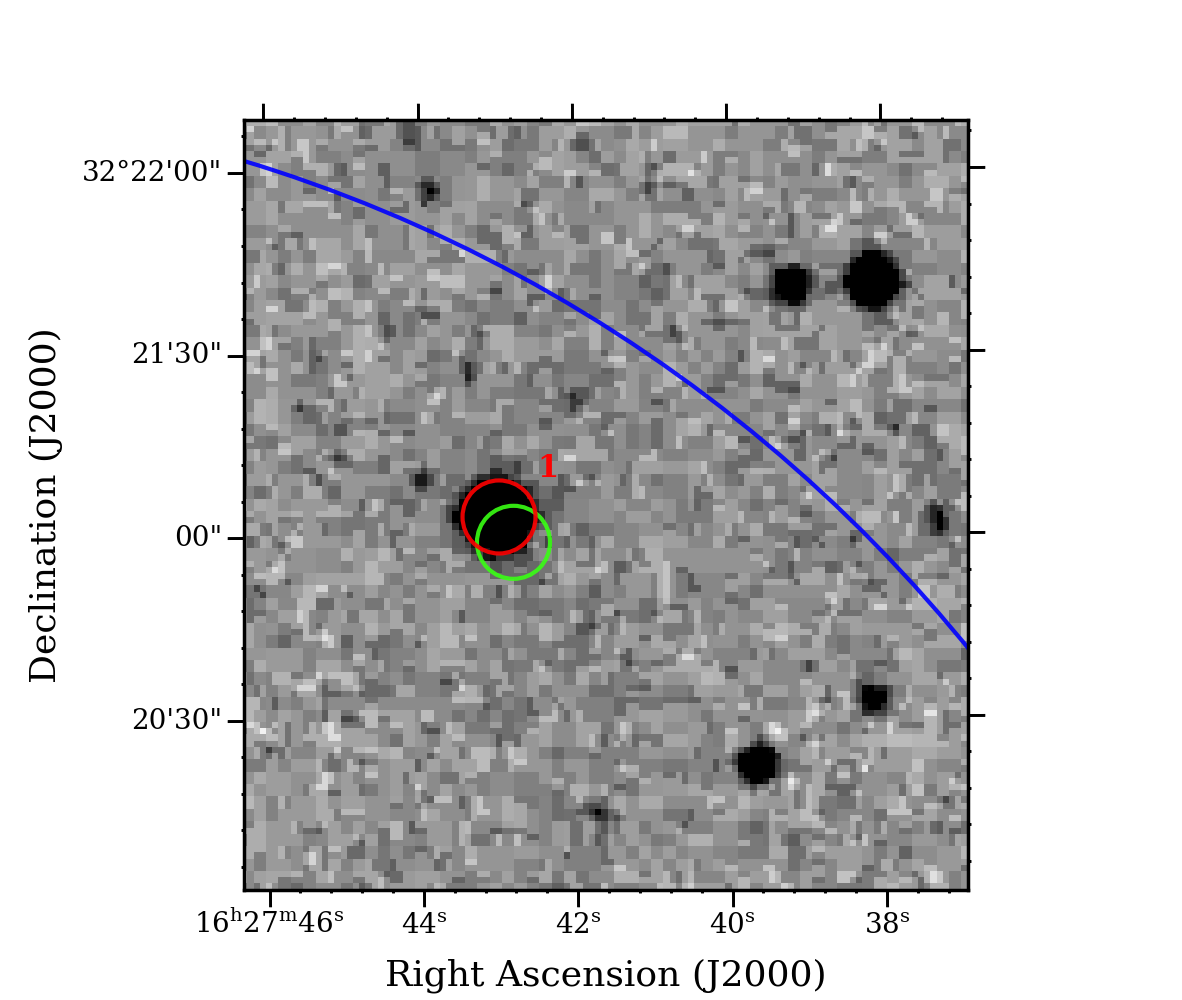}
    \caption{Field of the $\gamma$-ray source 3FGL J1627.8+3217, showing the Fermi error ellipse (blue), the X-ray candidate counterpart selected according to our method (green), the counterpart with an optical spectrum available in the literature (magenta), the optical counterparts located within the Swift positional uncertainty (red). All images were taken from DSS.\textbf{Left panel:} Full field of the gamma source. \textbf{Right panel:} Zoom-in on the region where the X-ray detection and the optical counterpart labeled as "1" can be seen.}
    \label{fig:J1627.8+3217}
\end{figure*}

\begin{figure}[h!]
    \centering
    \includegraphics[width=0.95\columnwidth]{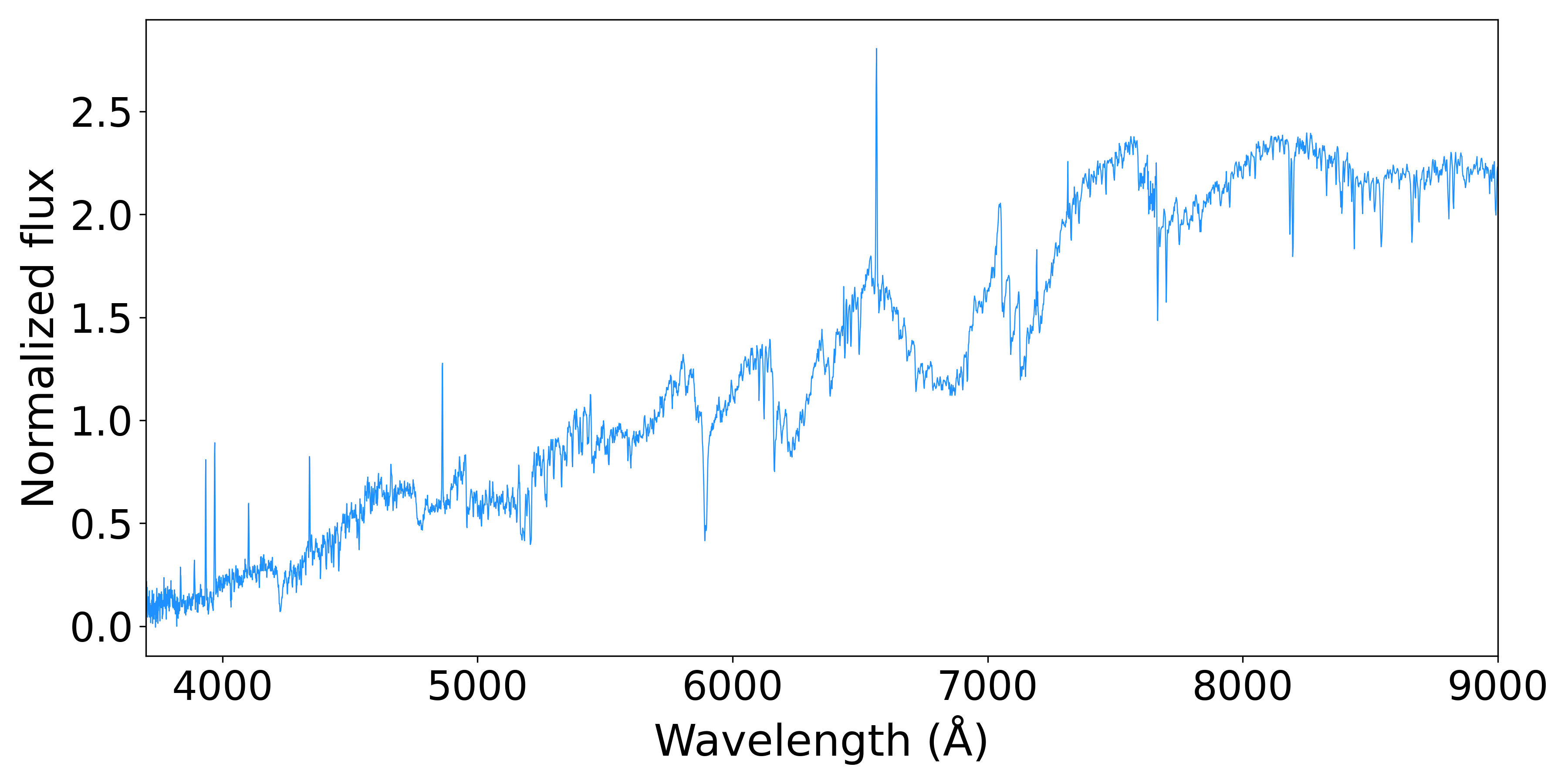}
    \caption{LAMOST spectrum of the optical counterpart of 3FGL J1627.8+3217 associated with the X-ray detection.}
    \label{fig:J1627.8+3217_spec}
\end{figure}

In this case, the optical counterpart previously proposed in the literature lies several arc minutes outside the Fermi error ellipse (see Fig.~\ref{fig:J1627.8+3217}). Conversely, our method yields an X-ray detection located within the confidence region, coinciding spatially with an object whose optical spectrum (see Fig.~\ref{fig:J1627.8+3217_spec}), as recorded in the LAMOST catalog, presently resembles that of a star (a red dwarf of M2 spectral type according to their classification). Despite this stellar classification, the source exhibits strong emission across the optical, ultraviolet, and X-ray bands, notably featuring a hard X-ray spectrum. Since no other viable candidate capable of driving the observed $\gamma$-ray emission was found inside the Fermi positional uncertainty region, the true nature of this source remains an open question.

\subsubsection{3FGL J2015.3-1431}\label{sec:J2015}

\begin{figure*}[t!] 
    \centering
    \includegraphics[width=0.49\textwidth]{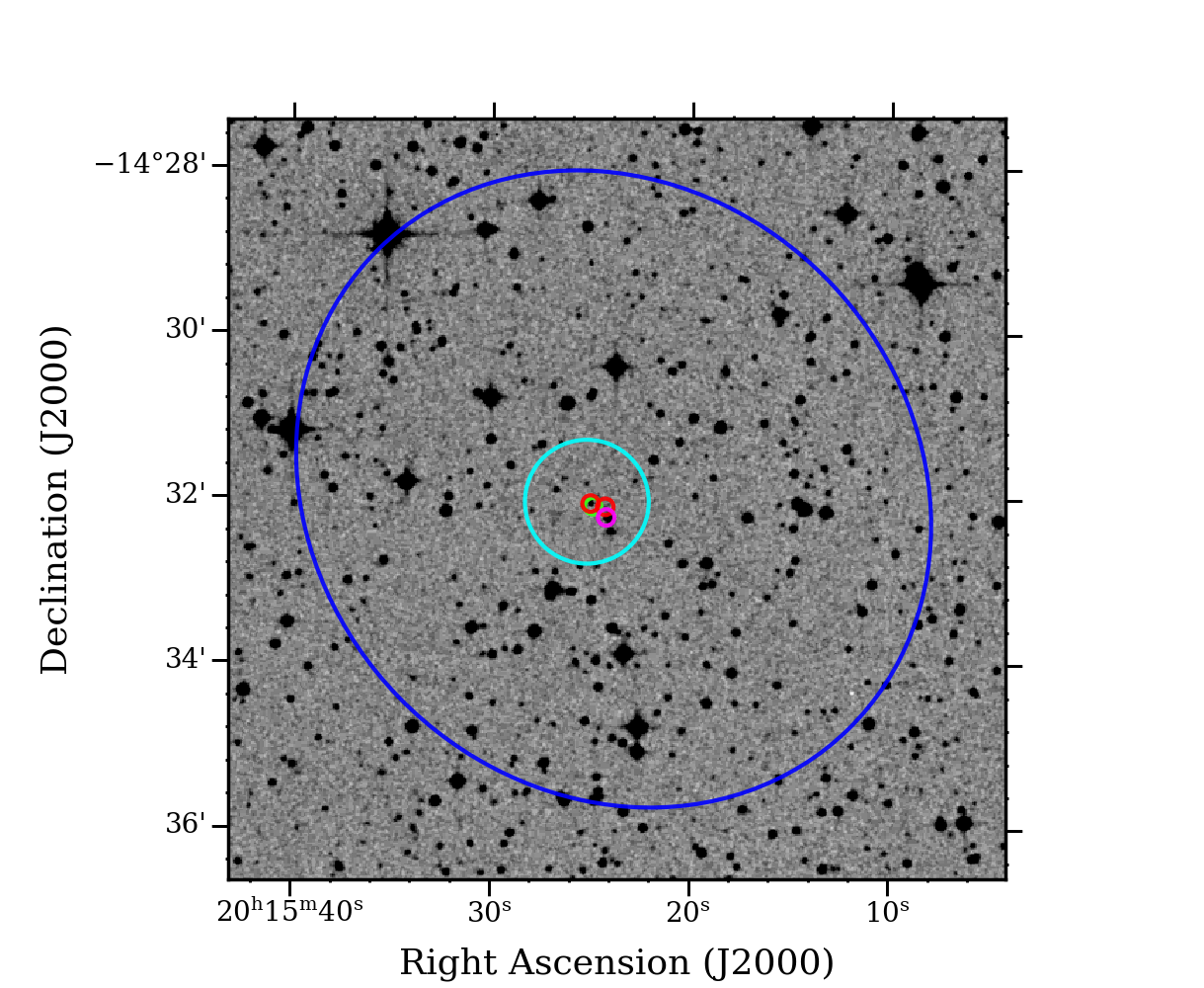}\hfill
    \includegraphics[width=0.51\textwidth]{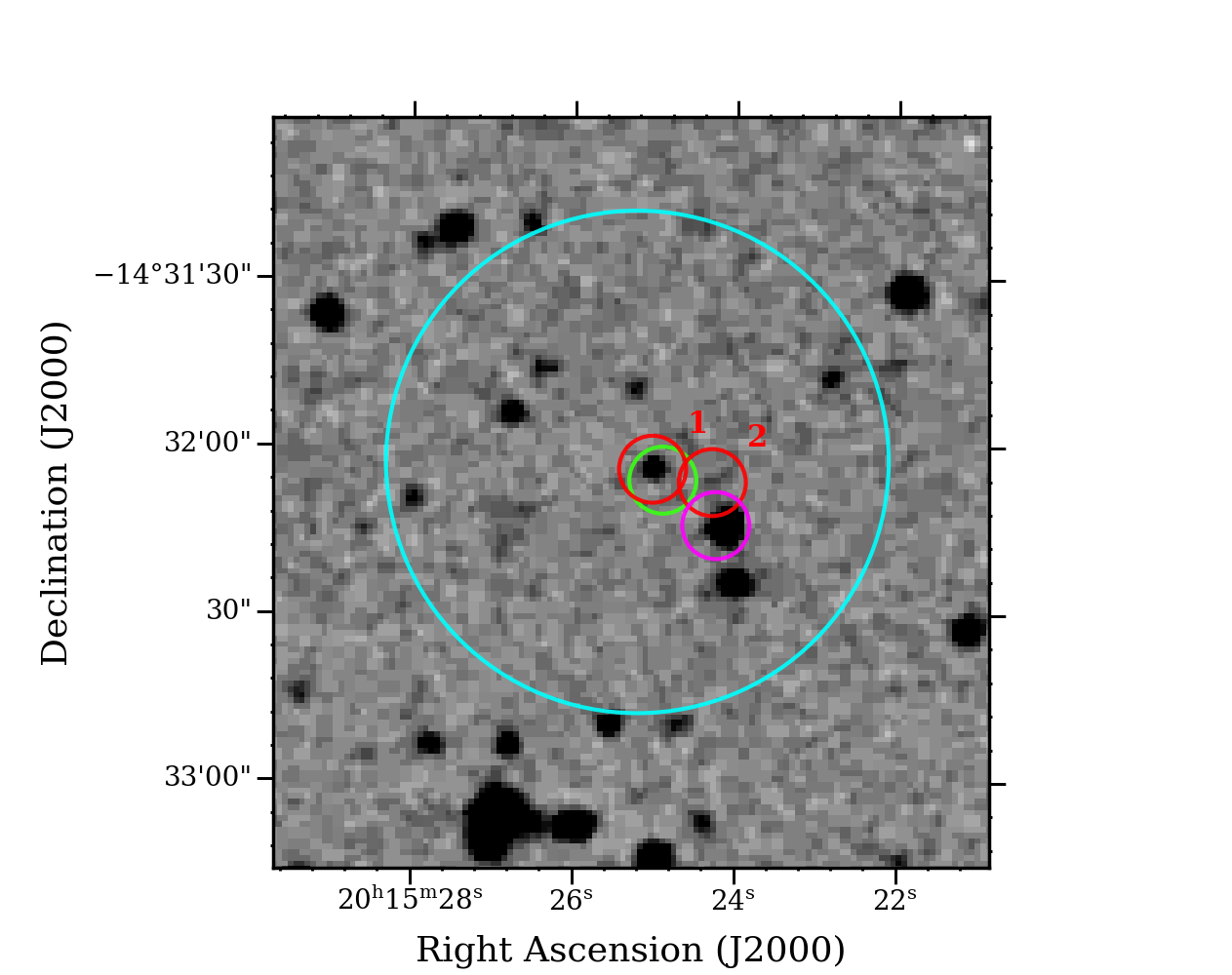}
    \caption{Field of the $\gamma$-ray source 3FGL J2015.3-1431, showingthe Fermi error ellipse (blue), the X-ray candidate counterpart selected according to our method (green), the counterpart with an optical spectrum available in the literature (magenta), the optical counterparts located within the Swift positional uncertainty (red), and the radio counterpart (NVSS, cyan). All images weretaken from DSS. \textbf{Left panel:} Full field of the gamma source. \textbf{Right panel:} Zoom-in on the region where two optical counterparts can be seen inside the Swift error.}
    \label{fig:J2015.3-1431}
\end{figure*}

The optical counterpart reported in the literature is located at an angular distance of 12.51 arc seconds from the proposed X-ray source (see Fig.~\ref{fig:J2015.3-1431}). Although this offset exceeds the standard 6 arc seconds confidence radius, the Swift-XRT positional uncertainty scales with the source counts as $\sim 23'' \times (\mathrm{counts})^{-0.48}$\citep{moretti2006}, implying larger uncertainties for low-count detections. Evaluating this expression for this case yields a positional uncertainty of 4.1 arc seconds. Even under this assumption, the reported optical source remains outside the derived error radius, suggesting that the spectroscopically identified counterpart is unlikely to be associated with the X-ray detection. Nevertheless, we identified a distinct optical source strictly coincident with the X-ray position, for which no optical spectrum is currently available. This object, Gaia DR3 6876244599401143552 (G=19.23 mag), exhibits infrared colors placing it within the blazar strip (W1-W2 = 0.475, W2-W3 = 1.763) and is also spatially associated with the radio source NVSS 201525-143202, which has a flux density of 13 mJy. These combined infrared colors and radio flux density strongly suggest that the source displays properties characteristic of blazars and will be observed in an upcoming follow-up campaign.

\subsection{Sources associated in the literature not spatially coincident with the X-ray counterparts}

Regarding the six sources associated in the literature but not spatially coincident with the X-ray counterparts identified, four are BL Lac objects, one is a QSO, and one is a FSRQ. All of them are classified as AGNs. For blazars (both BL Lacs and FSRQs), we cannot rule out that multiple sources may contribute to the $\gamma$-ray emission detected by Fermi. In the case of the QSO, while these objects are expected to be a major contaminant due to their abundance (e.g., \citealp{abdollahi}) and frequent detection in X-rays, they are not generally expected to be strong $\gamma$-ray emitters. 

Of those six sources, two have already been discussed in Sect.~\ref{sec:peculiar}: 3FGL J0006.2+0135 and 3FGL J2015.3-1431, in Subsections~\ref{sec:J0006} and \ref{sec:J2015}, respectively. For two other sources (3FGL J1228.4-0317 and 3FGL J1804.1+2532), the optical counterparts reported in the literature lie outside the 95\% Fermi error ellipse, so it is possible that they are not the true X-ray counterpart of these Fermi sources. Moreover, in the case of 3FGL J1228.4-0317, the proposed counterpart is classified as a QSO, for which $\gamma$-ray emission is not typically expected. 

As of 3FGL J0936.3-2114, the optical counterpart reported in the literature is associated with an X-ray source detected by Swift-XRT with $S/N > 3$. However, this source satisfies only one of the three criteria adopted from \cite{marchesini2019}. Although radio emission is detected, it does not meet the optical color criterion based on the $u-r$ color, and the W3 infrared measurement does not satisfy the required $S/N > 3$ threshold. This may suggest that the adopted criteria are too restrictive in some cases, potentially excluding plausible counterparts.

Finally, 3FGL J2043.6+0001 is classified as a BL Lac object in the literature. However, we find no X-ray counterpart associated with this source in the Swift-XRT observations. Since X-ray emission is generally expected in blazars as part of their broadband SED, we inspected the available Swift-XRT observations of this field, which amount to a total exposure time of approximately 3.5 ksec. Despite this coverage, no X-ray counterpart was detected. While this does not exclude a blazar nature, it places the source among the small fraction ($\sim$ 3\%) of blazars that remain undetected in X-rays with an exposure time over 1 ksec (e.g., \citealp{marchesini2020}). The reason for the lack of an X-ray detection in these objects is still not fully understood. Therefore, both this source and our proposed counterpart in the same Fermi field warrant further investigation and are planned targets of future follow-up observations. These cases further highlight the complexity of multiwavelength associations and motivate optical spectroscopic follow-up of the selected X-ray sources.

\section{Discussion}\label{sec:disc}

\begin{figure}[h!]
    \centering
    \includegraphics[width=0.95\columnwidth]{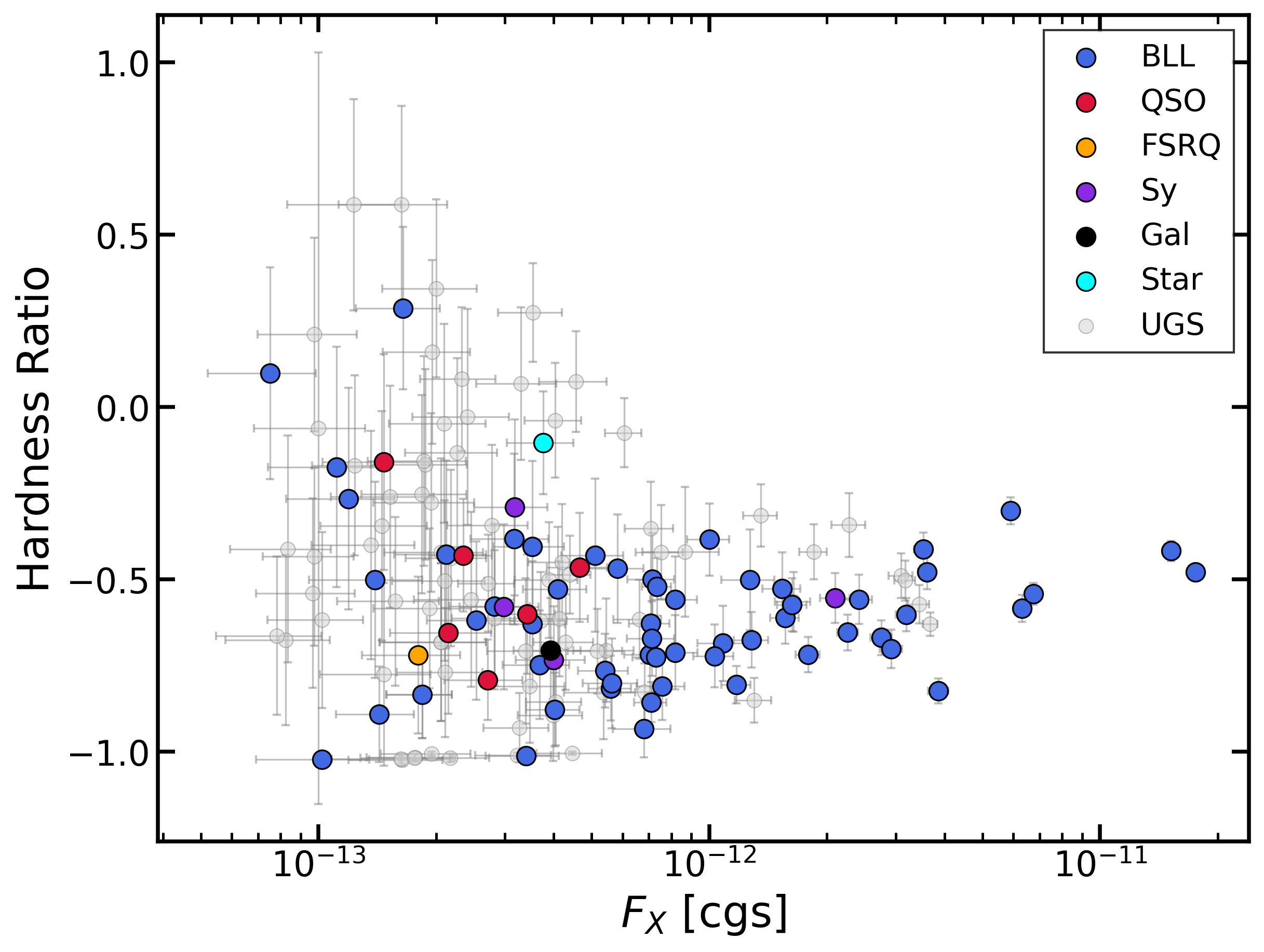}
    \caption{Hardness ratio vs. flux diagram for the whole sample. The different symbols indicate the different classes defined in this work.}
    \label{fig:HR_flux}
\end{figure}

In this work, we focused on the low-flux tail of the distribution in the HR versus flux plane identified in \cite{marchesini2020}, where the empirical separation becomes increasingly uncertain (see Fig.~\ref{fig:HR_flux}). The hardness ratio versus flux threshold depends on the current set of classified sources and is expected to evolve as new spectroscopic identifications are added. Expanding the number of confirmed sources is therefore essential to improve the statistical separation between populations and to assess possible contamination from non blazar objects. 

We investigated the location of the identified sources in the infrared color-color diagram by crossmatching our sample with the WISE catalog. Considering only sources with $S/N > 3$ in the three WISE bands, we obtain a subsample of 59 identified sources and 55 UGSs. Using the parameterization from \citep{massaro2012}, we find that most sources are consistent with the infrared classification scheme and lie within the blazar strip region \citep{massaro2011,dabrusco2012} which is associated with infrared emission dominated by nonthermal processes. Interestingly, two of the four identified Seyfert galaxies also fall within this region, suggesting that some $\gamma$-ray-emitting Seyferts may be associated with jetted active galaxies, although this scenario remains under debate \citep[e.g.,][]{paliya2019,foschini2020}.

\begin{figure}[h!]
    \centering
    \includegraphics[width=0.95\columnwidth]{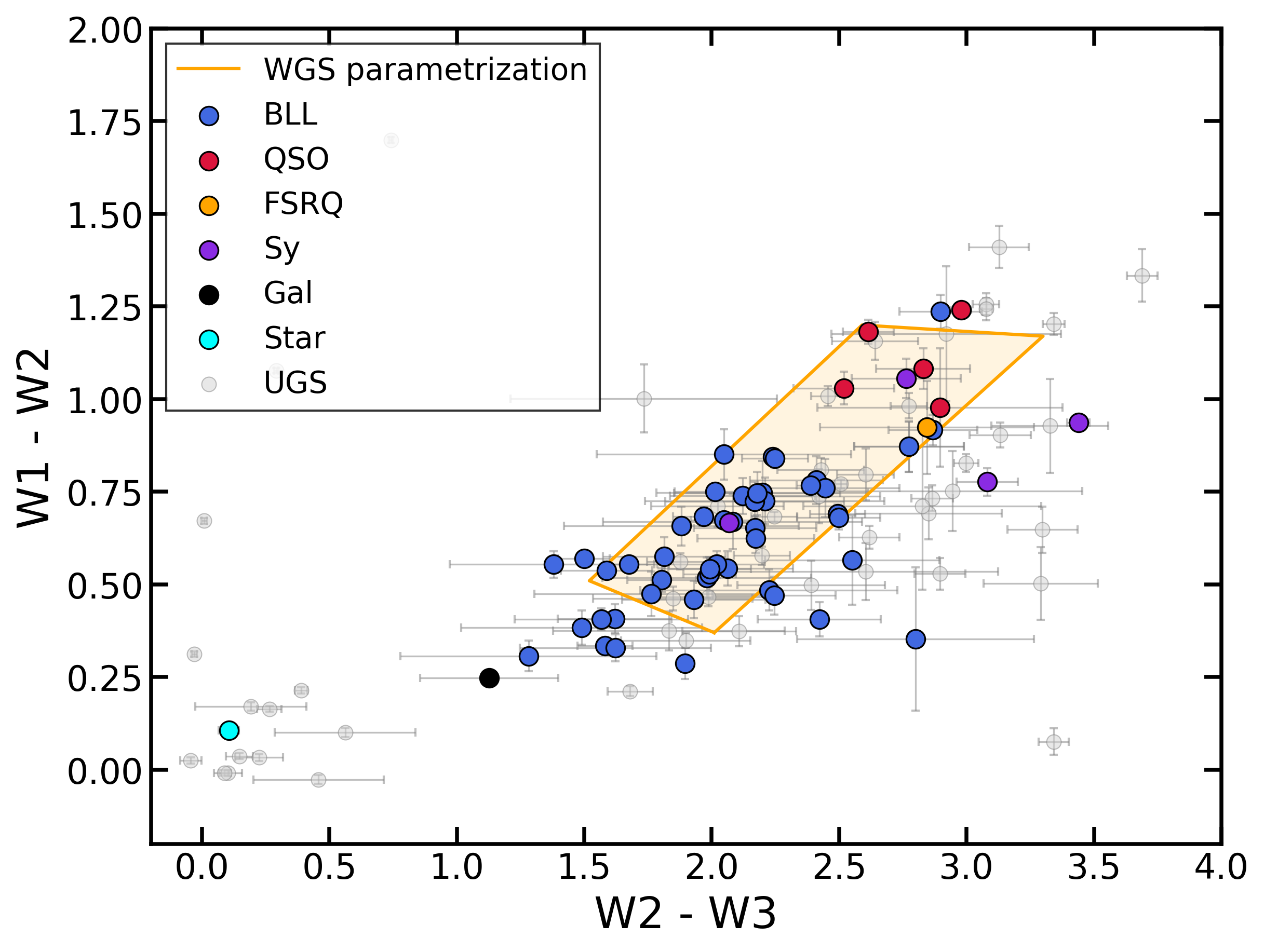}
    \caption{WISE color-color diagram for the sample. Sources are shown after applying a S/N $\geq$ 3 cut in the W1, W2, and W3 bands. The different symbols indicate the different classes defined in this work. The orange-shaded region and contour outline the WISE gamma-ray strip (WGS) blazar parameterization, as defined in \citep{massaro2012}.}
    \label{fig:WISE}
\end{figure}

\begin{figure*}[t!] 
    \centering
    \includegraphics[width=0.49\textwidth]{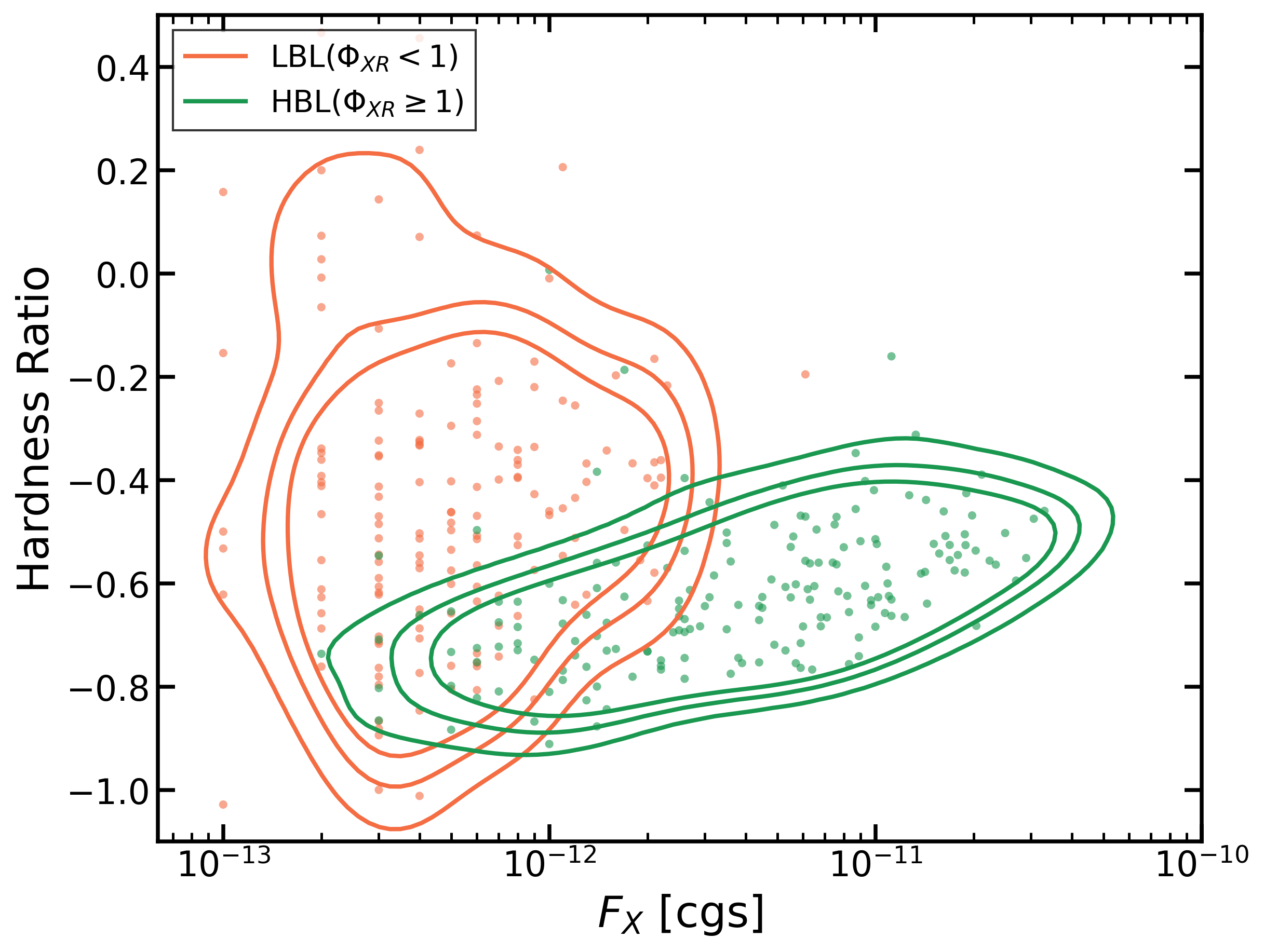}\hfill
    \includegraphics[width=0.49\textwidth]{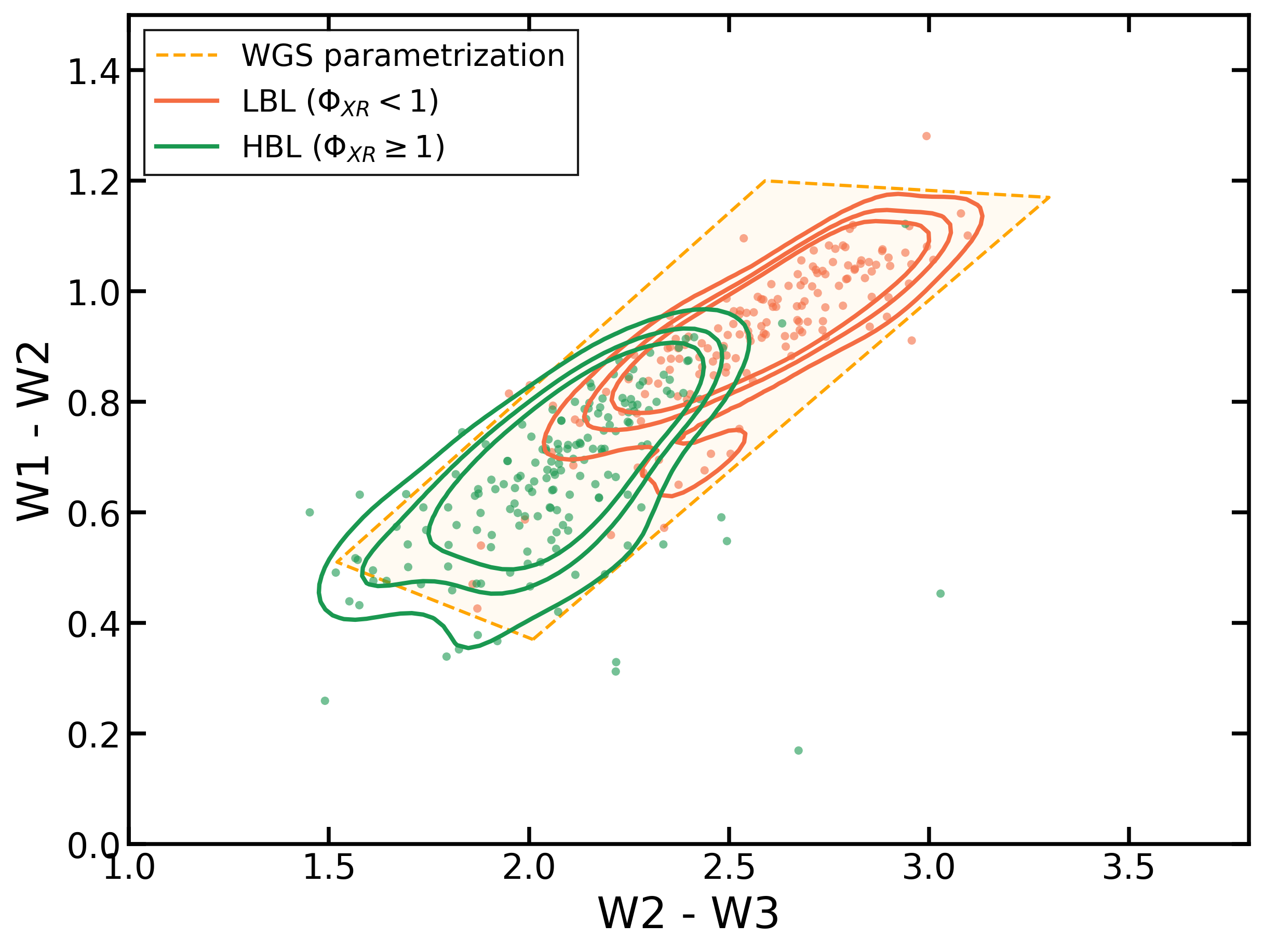}
    \caption{{70\%, 80\%, and 90\% isodensity contours of the integrated KDE distribution, as obtained from the KDE analysis on the whole BL Lacs Roma-BZCAT sample superimposed onto a characteristic diagram}. 
\textbf{Left panel:} Flux vs. $\mathrm{HR_x}$. \textbf{Right panel:} IR color-color diagram. We plot the LBL sources in orange and the HBL sources in green.} 
    \label{fig:contornos}
\end{figure*}

To distinguish between the HBL or LBL sources among the objects classified as BL Lacs, we adopted an approach similar to that in \cite{marchesini2016}. According to this, contour maps were constructed from 2D Gaussian kernel density estimations (KDEs) of $\gamma$-ray blazars drawn from the Roma-BZCAT catalog \citep{massaro2015}, in both the X-ray flux versus hardness ratio plane and the WISE color-color diagram. The parameter $\Phi_{XR}$ was then used as a diagnostic threshold (see Fig.~\ref{fig:contornos}). 

Building on these KDE density maps, we further used the underlying probability distributions to statistically predict the behavior of the sample sources. For this purpose, we evaluated the density of each source within the HBL and LBL distributions in both the X-ray plane (flux vs. hardness ratio) and the IR color-color diagram.

To estimate if a source belongs to the LBL or HBL populations, we interpreted the KDE as a relative likelihood. This allows us to estimate a class-dependent probability by comparing the density of the HBL and LBL distributions evaluated at the position of the particular source.

In the X-ray plane, this probability is defined as
\begin{equation}
P(\mathrm{HBL)_{X}} = \frac{\mathrm{KDE_{HBL,X}}}{\mathrm{KDE_{HBL,X}} + \mathrm{KDE_{
LBL,X}}},
\end{equation}
with an analogous definition for the IR color-color diagram, $\mathrm{P(HBL)_{IR}}$.

To combine both diagnostics, we used the arithmetic mean of the two estimates,

\begin{equation}
P_{{final}} = \frac{\left(P(\mathrm{HBL})_{X} + P(\mathrm{HBL})_{IR}\right)}{2}.
\end{equation}

This approach accounts for missing infrared data, in which case the classification relies solely on the X-ray-based probability. We then defined a threshold on the resulting ensemble probability to classify the sources. Objects with $P_{final} \ge 0.60$ were classified as HBL candidates, while those with $P_{final} \le 0.40$ were classified as LBL candidates. Sources in the intermediate range ($0.40 < P_{final} < 0.60$) were considered uncertain (UC),  either because they may represent intermediate cases or because the available data do not allow to strongly constrain their classification. This way, we identify extreme cases, while leaving ambiguous sources in the UC category.

\begin{figure*}[t!] 
    \centering
    \includegraphics[width=0.49\textwidth]{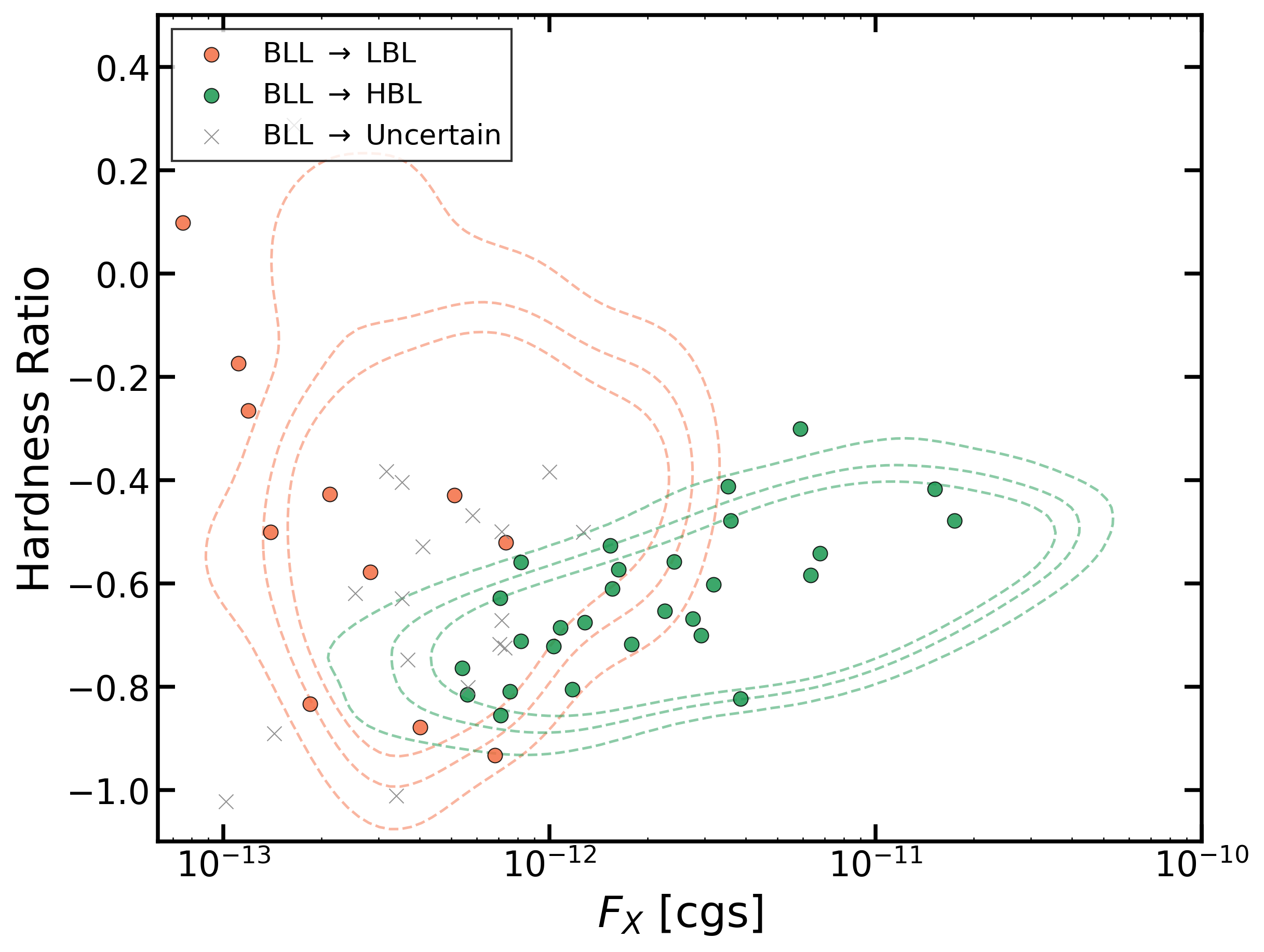}\hfill
    \includegraphics[width=0.49\textwidth]{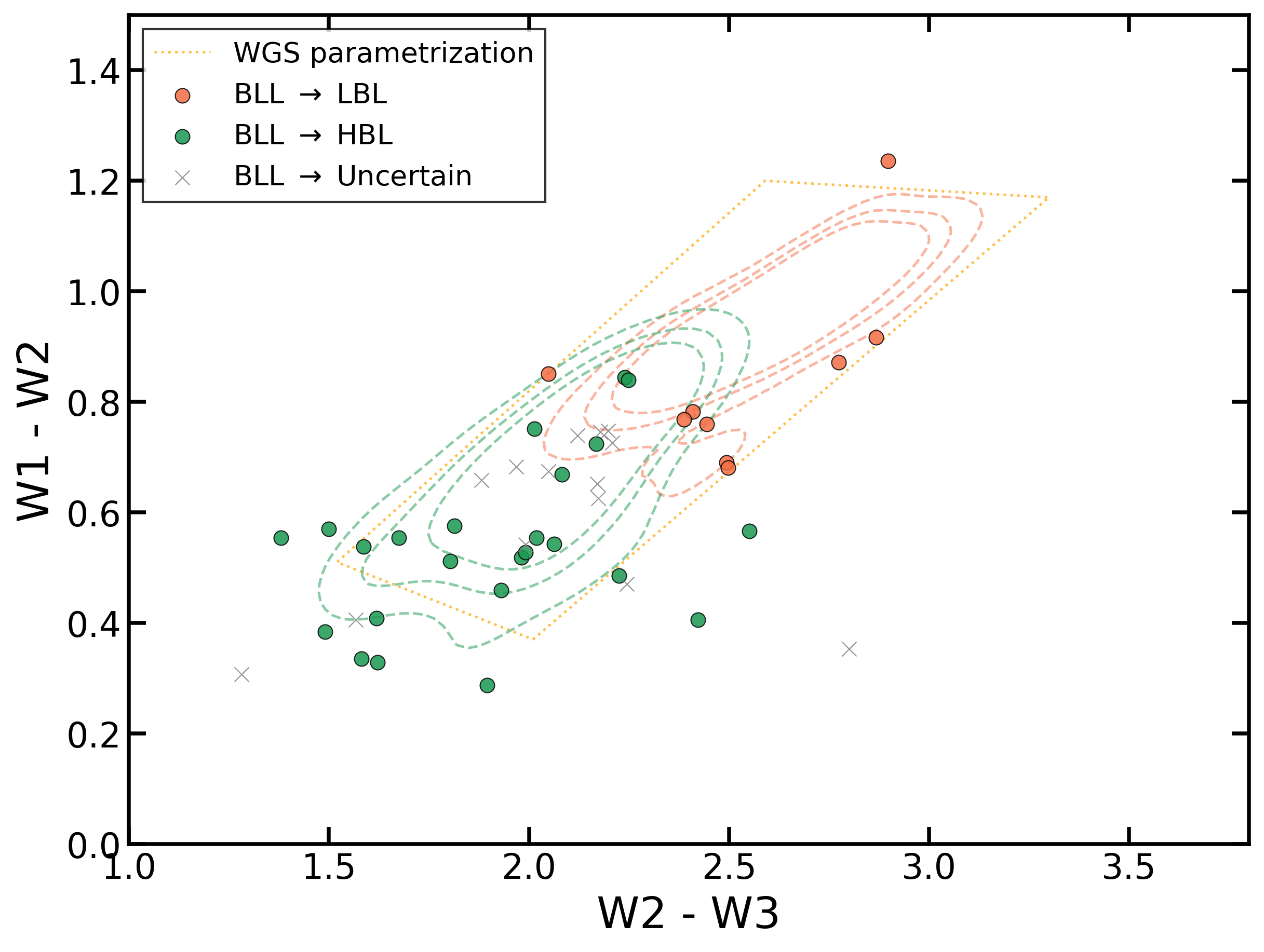}
    \caption{70\%, 80\%, and 90\% isodensity
contours, as obtained from the KDE analysis on the Roma-BZCAT sample and the BLL sample. The LBL candidate sources are shown in orange and the HBL candidate sources  in green. \textbf{Left panel:} Flux Vs. $\mathrm{HR_x}$. \textbf{Right panel:} Wise color-color diagram.} 
    \label{fig:contornosBLL}
\end{figure*}

\begin{figure*}[t!] 
    \centering
    \includegraphics[width=0.49\textwidth]{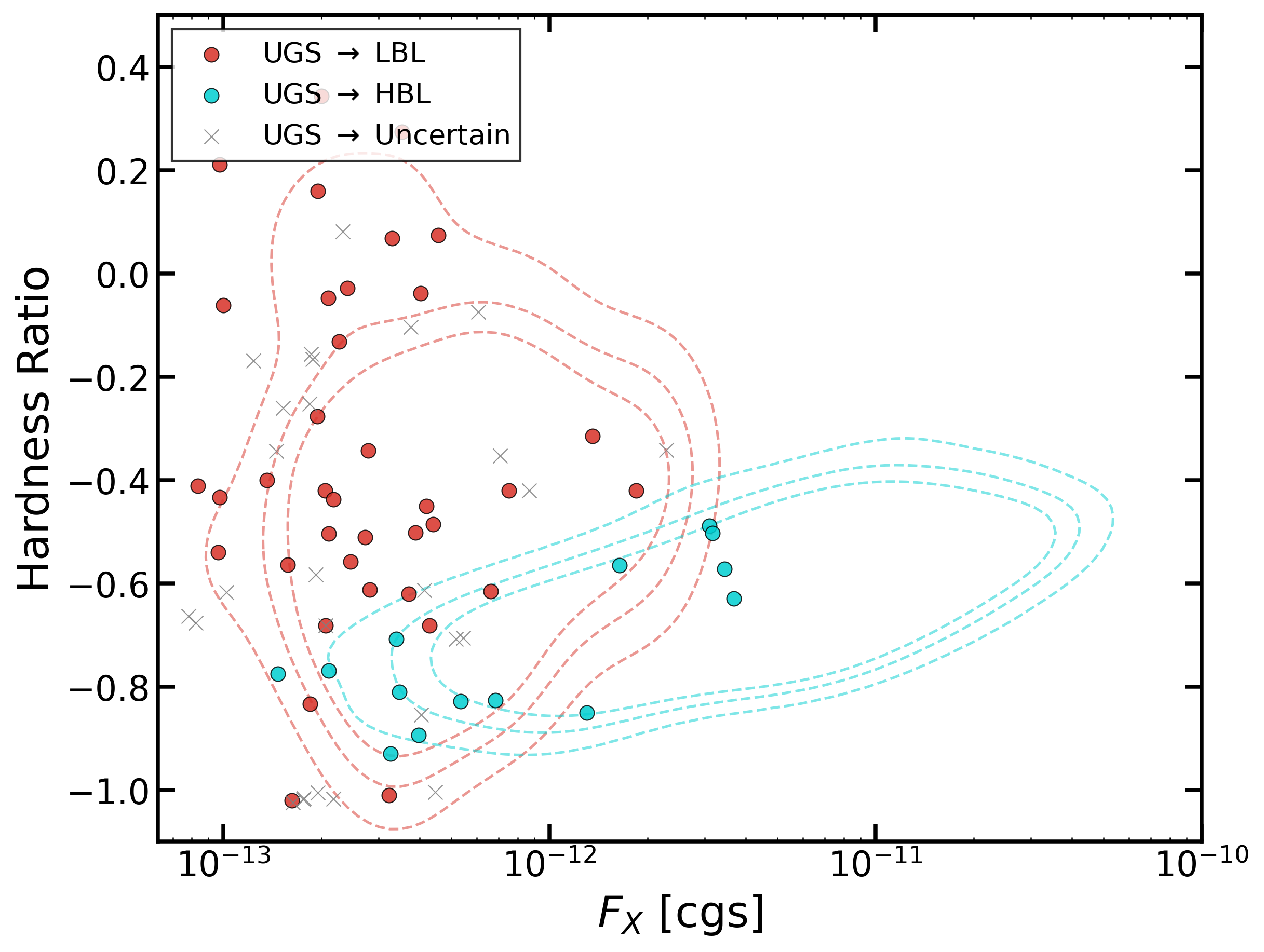}\hfill
    \includegraphics[width=0.49\textwidth]{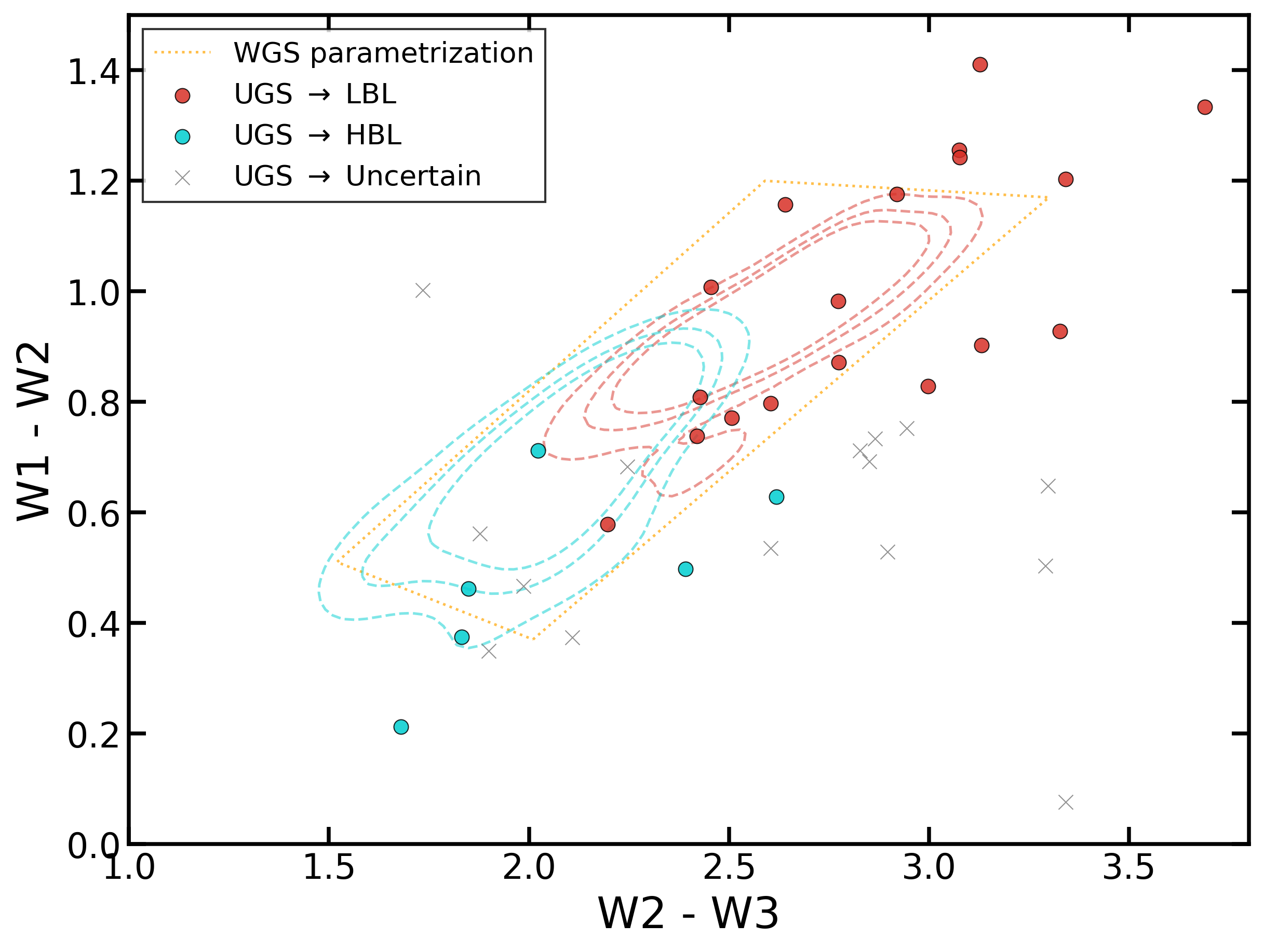}
    \caption{70\%, 80\%, and 90\% isodensity
contours, as obtained from the KDE analysis on the Roma-BZCAT sample and the UGS sample. The LBL candidate sources are shown in red and the HBL candidate sources  in cyan. Both correspond to UGS. \textbf{Left panel:} Flux vs. $\mathrm{HR_x}$. \textbf{Right panel:} Wise color-color diagram.} 
    \label{fig:contornosUGS}
\end{figure*}

We applied this procedure independently to the objects classified as BL Lac (see Fig.~\ref{fig:contornosBLL}) as well as to the UGS that could not be associated in this work due to the lack of available spectroscopic information in the literature (see Fig.~\ref{fig:contornosUGS}). This was done in order to obtain an indication of their nature. We find that out of the 57 BL Lac objects, 28 are HBL candidates, 11 are LBL candidates and the remaining 18 are UC. We compared our classification results with those reported by \cite{paiano2017} to assess their consistency and to evaluate whether our predictions are compatible with previous determinations (see Table. \ref{tab:comparacion}). Among the sources classified in both studies (excluding UC), we find a total of seven objects in common, and the results obtained in this study are consistent with theirs. Among the 80 UGS sources, 14 show characteristics compatible with the HBL classification and 39 with the LBL. Twenty-seven are considered UC. 

\begin{table}
\centering
\caption{List of sources in common between our sample and that in \cite{paiano2017}.}
\begin{tabular}{ccc}
\hline\noalign{\smallskip}
Name 3FGL & KDE & Literature \\
\hline
3FGLJ0049.0+4224 & UC & HBL \\
3FGLJ0239.0+2555 & HBL & HBL \\
3FGLJ0644.6+6035 & HBL & HBL \\
3FGLJ0937.9-1435 & UC & HBL \\
3FGLJ0952.8+0711 & UC & HBL \\
3FGLJ1049.7+1548 & HBL & HBL \\
3FGLJ1129.0+3758 & LBL & LBL \\
3FGLJ1411.4-0724 & HBL & IBL/HBL \\
3FGLJ1704.1+1234 & HBL & HBL \\
3FGLJ1704.4-0528 & HBL & HBL \\
3FGLJ2115.2+1215 & UC & HBL \\

\hline
\end{tabular}
\tablefoot{Column 1 lists the name of the source; column 2, the classification using the KDE criteria; and column 3, the classification according to \cite{paiano2017}.}
\label{tab:comparacion}
\end{table}

\begin{figure}[h!]
    \centering
    \includegraphics[width=0.95\columnwidth]{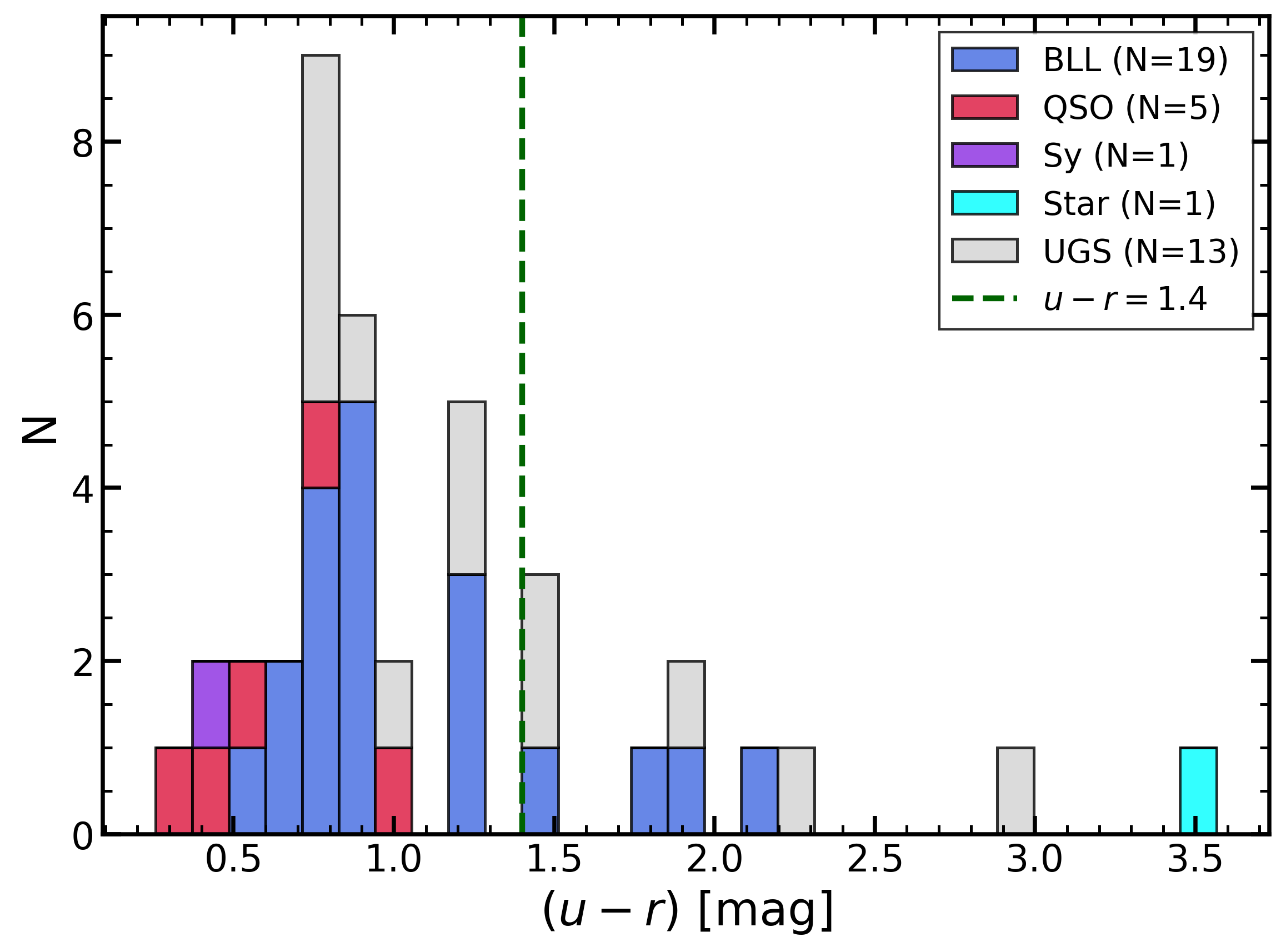}
    \caption{Histogram of the SDSS $u-r$ color index for the sources in our sample. The vertical line at $u-r$ = 1.4 mag represents the separation between BL Lac objects and other sources.}
    \label{fig:u-r}
\end{figure}

For completeness, we crossmatched our sample with the Sloan Digital Sky Survey Data Release 12 (SDSS DR12) photometric catalog to obtain the optical $u-r$ color indexes. According to \cite{massaro2012b}, an index value lower than 1.4\,mag is indicative (a proxy at optical bands) to the BL Lac nature of an object. The results can be seen in Fig.~\ref{fig:u-r}. Out of the 19 BL Lacs with available SDSS counterparts in the crossmatched sample, 15 (78.9\%) lie below the 1.4 mag threshold, consistent with BL Lac colors. One of the sources in this crossmatch was a Seyfert galaxy, with $u-r = 0.45$.

Regarding the Seyfert galaxies of our sample, two of the four lie within the infrared color space typically occupied by blazars in the WISE blazar strip (see Fig.~\ref{fig:WISE}). Additionally, the only Seyfert galaxy identified in the crossmatch with SDSS-DR12 exhibited a blazar-like $u-r$ color. This positional overlap may indicate that these sources are not classical Seyfert galaxies but could instead be misclassified as narrow-line Seyfert 1 (NLS1) objects. Narrow-line Seyfert 1 objects are a class of AGNs, of which only a small fraction are detected as $\gamma$-ray emitters ($\gamma$-NLS1s). They can exhibit blazar-like properties, often interpreted as systems hosting relativistic jets aligned close to the line of sight \cite{paliya2019}. In this context, $\gamma$-NLS1s have been proposed to represent the low black hole mass, high accretion rate end of the FSRQ population \cite{foschini2020}; therefore, it is not unexpected that this type of object may appear in samples aimed at identifying blazar candidates. 

The available radio information provides only limited support for this interpretation. Of the four Seyfert galaxies, only one (3FGL J2358.5+3827) exhibits clear radio emission, with a flux density of 57.4 mJy reported by NVSS. For two sources (3FGL J0031.6+0938 and 3FGL J1234.7-0437), only upper limits on the radio flux could be established from FIRST observations ($\sim$ 0.9 mJy), while no radio information was found in the available catalogs for the remaining object (3FGL J1306.8-4031).

With respect to the QSOs identified in our sample, their radio properties do not support a blazar classification. Indeed, all objects with available radio information are consistent with the RQ population, making it unlikely that they are misclassified FSRQs. This is consistent with their location in a region of parameter space where the distinction between QSOs and FSRQs is less ambiguous, and this reduces the likelihood of misclassification based on radio properties alone. FSRQs typically exhibit stronger radio emission due to their relativistic jets.

We also crossmatched our sample with the 4FGL-DR4 catalog to check for consistencies or discrepancies. The results were mostly consistent: of the 57 objects identified as BL Lacs in our sample, 45 were associated with a source in the 4FGL-DR4. Of these, 40 were classified as BL Lacs, four as BCUs (blazar candidates of unknown types), and one as an FSRQ.  While an in-depth analysis of the 4FGL-DR4 results is beyond the aims of this work, these cases will be investigated in a future work (Weiss, J. L. et al. in prep).

\section{Summary and conclusions}

In the present work, we sought to empirically validate the capability of the method proposed by \cite{marchesini2019} and \cite{marchesini2020} to classify Fermi sources based on X-ray properties. Starting with a sample of 148 sources counting at least one X-ray detection within their positional error ellipse, a cross-match was performed using the optical catalogs Simbad and Aladin, and an exhaustive literature search was conducted to verify the existence of optical spectra for these sources. This was further complemented by searches for sources not found in the literature through the correct identification of optical spectra in SDSS-DR19 and the subsequent search for radio counterparts. Following this, a subsample was defined, including those sources associated in the literature with the Fermi source but not spatially compatible with the optical source selected by the X-ray criteria.

It should be noted that the Fermi-LAT 4FGL-DR4 catalog lists 4635 associated or identified sources, of which 32.15\% are BL Lac blazars. Many of the still unclassified sources remain unassociated due to being too faint or having large positional uncertainties. It is expected that the percentages over the unclassified objects in each class are similar. Therefore, when analyzing a sample of unclassified or unassociated sources, a successful classification method should recover a proportion of BL Lacs equal to or higher than 32.15\%. Another important aspect is that the 95\% Fermi error ellipses represents a statistical limit for any proposed classification method, as approximately 1 in 20 Fermi detections is expected to have its actual counterpart lying outside this region. In addition, the association process is affected by the possibility of false positives. Using the 2SXPS Swift-XRT point source catalog \citep{Evans20}, and applying the same selection criteria adopted in this work, we estimated a background X-ray source density of $\rho \sim 2.81 \,\mathrm{deg}^{-2}$. By combining this with the total area of the 95\% confidence Fermi error ellipses in our sample, we derive an expected number of spurious associations of $N_{\mathrm{FP}} \sim 4 \pm 2$, corresponding to approximately 6\% of the sample. 

The X-ray-based identification criterion has proven particularly effective in recognizing BL Lacs, suggesting that a significant fraction of the still unassociated sources in the Fermi-LAT catalog could belong to this class. This is supported by the obtained IF, which reached 2.50 when considering optical sources within the Swift positional uncertainty. Moreover, the method enables source classification without requiring optical spectroscopy, making it an efficient alternative for large samples. The predominance of BL Lacs among the identified UGSs is consistent with population synthesis models of the $\gamma$-ray sky \citep[e.g.,][]{ackermann2015, ajello2020}, and our analysis confirms that most high-probability BL Lac candidates selected by \cite{marchesini2019} are indeed blazars.

The high recovery rate of BL Lacs highlights the close connection between X-ray and $\gamma$-ray emission in these sources. Since both bands probe the most energetic regions of relativistic jets, X-ray properties provide reliable indicators of jet dominance and an efficient means of distinguishing blazars from the broader AGN population.

We further explored the nature of the BL Lac population by combining the X-ray and infrared parameter spaces through a KDE based probabilistic classification. Applying this method to the confirmed BL Lac sample, we identified 28 HBL candidates and 10 LBL candidates, while 17 sources remained uncertain. A comparison with the classifications reported by \cite{paiano2017} showed full agreement for all sources in common, supporting the consistency of the adopted approach. The same methodology was also applied to the UGS sample, providing tentative indications of the nature of sources lacking spectroscopic classifications.

Moreover, the employed method shows an unexpected capability for detecting $\gamma$-ray-emitting Seyfert galaxies, whose emission mechanisms at these energies are still poorly understood. The performance of the method is validated by the obtained IF of $\sim22$. Such a significant increase in the IF  points out the high sensitivity of our approach toward this class of sources. Notably, the 4FGL-DR4 catalog lists a total of 12 associated or identified Seyfert galaxies; therefore, using this method, it is possible to increase the number of this particular type of source by more than 33\% of the total detected by Fermi. Other authors have reported similarly limited samples of these objects, for example, \cite{paliya2019} identified a total of 22 such sources, while \cite{foschini2022} reported 16. The small number of Seyferts detected at high energies known  highlights the extreme rarity of this population. Their $\gamma$-ray detection is very uncommon, making every confirmed $\gamma$-ray association a remarkable finding. However, the small number of detected sources prevents us from drawing definitive conclusions regarding the method efficiency for Seyferts, but this preliminary result is significant enough to motivate dedicated spectroscopic follow up campaigns. Testing whether this apparent excess of Seyfert galaxies persists in larger samples will be an important follow up of this work. Interestingly, two of the Seyfert galaxies identified in our sample lie within the WISE blazar strip, suggesting that they may belong to the population of $\gamma$-ray-emitting NLS1s rather than classical Seyferts. Furthermore, the only Seyfert galaxy with reported magnitudes in both the u and r bands exhibits a $u-r$ color consistent with those typically observed in BL Lac objects. Although the sample is too small to draw firm conclusions, given that the sources were identified among Fermi objects using a method designed to select BL Lac candidates, some degree of similarity in the flux properties used by the method is expected between BL Lacs and the $\gamma$-ray-emitting Seyfert galaxies found in our sample.

The results motivate spectroscopic follow-up of UGSs that still lack optical counterparts, both to expand the sample and to test whether the apparent excess of Seyfert galaxies persists. In particular, obtaining optical spectra for the high-probability blazar candidates identified by \cite{marchesini2019} would provide a direct validation of the proposed method.

Overall, our results confirm that X-ray data play a key role in linking optical and $\gamma$-ray observations for blazar identification. Applying this method to future Fermi-LAT catalogs will enable its validation on larger samples of unassociated sources, improving the blazar census and our understanding of the $\gamma$-ray sky.

\begin{acknowledgements}

Funding for the Sloan Digital Sky Survey V has been provided by the Alfred P. Sloan Foundation, the Heising-Simons Foundation, the National Science Foundation, and the Participating Institutions. SDSS acknowledges support and resources from the Center for High-Performance Computing at the University of Utah. SDSS telescopes are located at Apache Point Observatory, funded by the Astrophysical Research Consortium and operated by New Mexico State University, and at Las Campanas Observatory, operated by the Carnegie Institution for Science. The SDSS web site is \url{www.sdss.org}. SDSS is managed by the Astrophysical Research Consortium for the Participating Institutions of the SDSS Collaboration.

We acknowledge the use of public data from the Swift data archive. This publication makes use of data products from the Wide-field Infrared Survey Explorer, which is a joint project of the University of California, Los Angeles, and the Jet Propulsion Laboratory/California Institute of Technology, funded by the National Aeronautics and Space Administration. Part of this work is based on the NVSS (NRAO VLA Sky Survey). The National Radio Astronomy Observatory is operated by Associated Universities, Inc., under contract with the National Science Foundation. The Molonglo Observatory site manager, Duncan Campbell-Wilson, and the staff, Jeff Webb, Michael White, and
John Barry, are responsible for the smooth operation of the Molonglo Observatory Synthesis Telescope (MOST) and the day-to-day observing program of SUMSS. SUMSS is dedicated to Michael Large, whose expertise and vision made the project possible. The MOST is operated by the School of Physics with the support of the Australian Research Council and the Science Foundation for Physics within the University of Sydney. Guoshoujing Telescope (the Large Sky Area Multi-Object Fiber Spectroscopic Telescope LAMOST) is a National Major Scientific Project built by the Chinese Academy of Sciences. Funding for the project has been provided by the National Development and Reform Commission. LAMOST is operated and managed by the National Astronomical Observatories, Chinese Academy of Sciences. The present work was partially supported by grant 11/G178 from the Universidad Nacional de La Plata, Argentina. E. J. M. would like to thank Dr. R. Campana for his useful insights

\end{acknowledgements}

\bibliographystyle{aa}
\bibliography{bibliografia}

\clearpage

\begin{appendix}
\onecolumn
\section{Supplementary figures and tables}

\begin{longtable}{lcccccc}
\caption{UGS sources with optical spectra reported in the literature.}\label{tab:ugs} \\
\toprule
Name & R.A.$_x$ (J2000) & DEC$_x$ (J2000) & Class & Ref. & Comments \\

\midrule
\endfirsthead
\caption[]{UGS sources with optical spectra reported in the literature.} \\
\toprule
Name & R.A.$_x$ (J2000) & DEC$_x$ (J2000) & Class & Ref. & Comments \\

\midrule
\endhead
\midrule
\multicolumn{7}{r}{Continued on next page} \\
\midrule
\endfoot
\bottomrule
\endlastfoot
3FGLJ0006.2+0135 & 00:06:06.0 & +01:31:21.4 & BLL & 1  &  See Sect. \ref{sec:J0006}\\
3FGLJ0031.6+0938 & 00:31:59.7 & +09:36:17.1 & Sy & 1  &  \\
3FGLJ0049.0+4224 & 00:48:59.1 & +42:23:48.4 & BLL & 2 &  \\
3FGLJ0121.8-3917 & 01:21:52.5 & -39:15:44.2 & BLL & 3 &  \\
3FGLJ0138.5-4612 & 01:38:33.6 & -46:14:16.1 & BLL & 4 &  \\
3FGLJ0156.5-2423 & 01:56:24.4 & -24:20:03.9 & BLL & 3 &  \\
3FGLJ0158.6+0102 & 01:58:52.6 & +01:01:30.4 & BLL & 1  &  \\
3FGLJ0200.3-4108 & 02:00:20.7 & -41:09:34.9 & BLL & 3 &  \\
3FGLJ0234.2-0629 & 02:34:10.5 & -06:28:29.5 & BLL & 1  &  \\
3FGLJ0239.0+2555 & 02:38:54.1 & +25:54:07.3 & BLL & 2 &  \\
3FGLJ0251.1-1829 & 02:51:11.5 & -18:31:10.3 & BLL & 1  &  \\
3FGLJ0258.9+0552 & 02:58:57.5 & +05:52:43.2 & BLL & 1  &  \\
3FGLJ0340.4-2423 & 03:40:50.0 & -24:22:55.3 & QSO & 3 &  \\
3FGLJ0414.9-0840 & 04:14:33.0 & -08:42:07.5 & BLL & 1  &  \\
3FGLJ0420.4-6013 & 04:20:11.0 & -60:15:05.3 & BLL & 3 &  \\
3FGLJ0430.1-3103 & 04:29:58.8 & -30:59:34.9 & BLL & 4 &  \\
3FGLJ0437.7-7330 & 04:38:37.0 & -73:29:20.2 & Gal & 3 &  \\
3FGLJ0644.6+6035 & 06:44:36.1 & +60:38:52.9 & BLL & 2 &  \\
3FGLJ0704.3-4828 & 07:04:21.6 & -48:26:45.8 & BLL & 3 &  \\
3FGLJ0747.5-4927 & 07:47:24.8 & -49:26:32.6 & BLL & 3 &  \\
3FGLJ0826.3-6400 & 08:26:28.1 & -64:04:15.7 & BLL & 3 &  \\
3FGLJ0935.1-1736 & 09:35:14.9 & -17:36:55.7 & BLL & 5 &  \\
3FGLJ0936.3-2114 & 09:36:16.6 & -21:16:39.4 & BLL & 5 & Positional mismatch with \\
                 &            &             &     &   &  X-ray method source  \\
3FGLJ0937.9-1435 & 09:37:54.7 & -14:33:49.7 & BLL & 2 &  \\
3FGLJ0952.8+0711 & 09:52:49.6 & +07:13:31.2 & BLL & 2 &  \\
3FGLJ1016.6-4244 & 10:16:20.7 & -42:47:22.1 & BLL & 6 &  \\
3FGLJ1049.7+1548 & 10:49:39.4 & +15:48:39.0 & BLL & 2 &  \\
3FGLJ1052.0+0816 & 10:52:24.5 & +08:14:08.7 & BLL & 4 &  \\
3FGLJ1117.7+0217 & 11:17:31.0 & +02:15:55.0 & QSO & 7 &  \\
3FGLJ1120.6+0713 & 11:20:42.5 & +07:13:12.9 & QSO & 8 &  \\
3FGLJ1129.0+3758 & 11:29:03.6 & +37:56:58.7 & BLL & 2 &  \\
3FGLJ1130.7-7800 & 11:30:32.3 & -78:01:06.0 & BLL & 9 &  \\
3FGLJ1132.0-4736 & 11:32:09.4 & -47:38:53.5 & BLL & 3 &  See Sect. \ref{sec:J1132} \\
3FGLJ1146.1-0640 & 11:46:00.9 & -06:38:51.7 & BLL & 1  &  \\
3FGLJ1223.3+0818 & 12:23:27.4 & +08:20:30.9 & BLL & 1  &  \\
3FGLJ1225.4-3448 & 12:25:30.9 & -34:46:08.8 & BLL & 10 &  \\
3FGLJ1228.4-0317 & 12:28:19.5 & -03:18:12.8 & QSO & 8 &  Optical source outside the \\
 &  &  &  &  & Fermi 95\% error ellipse \\
3FGLJ1234.7-0437 & 12:34:47.9 & -04:32:47.4 & Sy & 1  &  \\
3FGLJ1249.5-0546 & 12:49:19.6 & -05:45:39.5 & BLL & 11 &  \\
3FGLJ1258.4+2123 & 12:58:21.5 & +21:23:51.7 & BLL & 1  &  \\
3FGLJ1301.5+3333 & 13:01:26.0 & +33:39:20.5 & FSRQ & 12 &  \\
3FGLJ1306.8-4031 & 13:06:56.2 & -40:35:22.0 & Sy & 13 &  \\
3FGLJ1309.0+0347 & 13:08:32.3 & +03:44:09.8 & QSO & 6 &  \\
3FGLJ1315.7-0732 & 13:15:53.0 & -07:33:01.5 & BLL & 14 &  \\
3FGLJ1410.9+7406 & 14:10:45.4 & +74:05:11.1 & BLL & 14 &  \\
3FGLJ1411.4-0724 & 14:11:33.3 & -07:22:54.4 & BLL & 2 &  \\
3FGLJ1517.0+2637 & 15:16:50.0 & +26:36:35.6 & QSO & 15 &  \\
3FGLJ1541.6+1414 & 15:41:50.1 & +14:14:42.2 & BLL & 1  &  \\
3FGLJ1549.9-3044 & 15:49:46.5 & -30:45:01.6 & BLL & 10 &  \\
3FGLJ1627.8+3217 & 16:27:42.8 & +32:20:59.0 & Star & 5 & See Sect. \ref{sec:J1627} \\
3FGLJ1704.1+1234 & 17:04:09.5 & +12:34:22.9 & BLL & 2 &  \\
3FGLJ1704.4-0528 & 17:04:33.8 & -05:28:39.6 & BLL & 2 &  \\
3FGLJ1804.1+2532 & 18:04:34.0 & +25:31:01.4 & FSRQ & 16 & Optical source outside the \\
 &  &  &  &  & Fermi 95\% error ellipse \\
3FGLJ1810.7+5335 & 18:10:37.9 & +53:35:01.8 & BLL & 5 &  \\
3FGLJ1923.2-7452 & 19:22:42.0 & -74:53:54.7 & BLL & 5 &  \\
3FGLJ1934.2+6002 & 19:34:19.8 & +60:01:40.7 & BLL & 11 &  \\
3FGLJ2015.3-1431 & 20:15:24.9 & -14:32:06.0 & BLL & 16 & See Sect. \ref{sec:J2015} \\
3FGLJ2030.5-1439 & 20:30:28.0 & -14:39:21.2 & QSO & 3 &  \\
3FGLJ2034.6-4202 & 20:34:50.9 & -42:00:37.7 & BLL & 9 &  \\
3FGLJ2043.6+0001 & 20:43:42.3 & +00:01:18.7 & BLL & 5 & Positional mismatch with \\
                 &            &             &     &   &  X-ray method source  \\
3FGLJ2110.0+0442 & 21:09:40.1 & +04:39:59.4 & BLL & 17 &  \\
3FGLJ2115.2+1215 & 21:15:22.1 & +12:18:01.9 & BLL & 2 &  \\
3FGLJ2144.6-5640 & 21:44:29.5 & -56:38:47.9 & BLL & 3 &  \\
3FGLJ2150.5-1754 & 21:50:46.6 & -17:49:56.3 & BLL & 1  &  \\
3FGLJ2209.8-0450 & 22:09:41.6 & -04:51:09.2 & BLL & 3 &  \\
3FGLJ2212.5+0703 & 22:12:31.0 & +07:06:51.6 & QSO & 1  &  \\
3FGLJ2229.1+2255 & 22:29:11.1 & +22:54:56.6 & BLL & 1  &  \\
3FGLJ2244.6+2503 & 22:44:36.9 & +25:03:42.9 & BLL & 3 &  \\
3FGLJ2247.2-0004 & 22:47:10.3 & -00:05:05.0 & BLL & 18 &  \\
3FGLJ2250.3+1747 & 22:50:32.8 & +17:49:14.7 & BLL & 1  &  \\
3FGLJ2258.2-3645 & 22:58:14.8 & -36:44:29.7 & BLL & 19  &  \\
3FGLJ2300.0+4053 & 23:00:12.3 & +40:52:22.6 & BLL & 17 &  \\
3FGLJ2321.6-1619 & 23:21:37.0 & -16:19:26.0 & BLL & 1  &  \\
3FGLJ2337.2-8425 & 23:36:24.1 & -84:26:50.4 & BLL & 3 &  \\
3FGLJ2351.9-7601 & 23:51:16.0 & -76:00:14.9 & BLL & 3 &  \\
3FGLJ2358.5+3827 & 23:58:25.2 & +38:28:58.0 & Sy & 1  &  \\
3FGLJ2358.6-1809 & 23:58:36.8 & -18:07:17.4 & BLL & 1  &  \\

\end{longtable}
\tablebib{
(1)~\citet{paiano2019}; (2) \citet{paiano2017}; (3) \citet{pena2017}; (4) \citet{pena2021};
(5) \citet{pena2020}; (6) \citet{ulgiati}; (7) \citet{ahumada}; (8) \citet{sexton};
(9) \citet{desai2019}; (10) \citet{rajagopal}; (11) \citet{garcia2023}; (12) \citet{massaro2016}; (13) \citet{linares}; (14) \citet{marchesini2016}; (15) \citet{rakshit}; (16) \citet{alvarez}; (17) \citet{marchesi2018} ; (18) \citet{shaw}; (19) \citet{landoni}.
}
\tablefoot{The columns show: (1) the Fermi source name; (2–3) the RA and Dec of the Swift X-ray counterpart; (4) the source class assigned in this work; (5) references; and (6) comments.}
\end{appendix}

\end{document}